\documentclass[journal,onecolumn]{IEEEtran}

\usepackage[dvips]{graphicx}
\usepackage{epsfig}
\usepackage[cmex10]{amsmath}
\usepackage{amssymb}
\usepackage{amsthm}
\usepackage{amsfonts}
\usepackage[pagebackref=false]{hyperref}
\usepackage{bm}
\usepackage{cite}
\usepackage[svgnames]{xcolor} 
\usepackage{pstricks,pst-node,pst-plot,pstricks-add}
\usepackage[tight,footnotesize]{subfigure}
\usepackage{algpseudocode}
\usepackage{algorithm}

\graphicspath{{figs/}}

\input{Definition.def}
\newcommand{\tabref}[1]{Table~\ref{#1}}
\begin{document}

\date{}

\title{CodedSketch: A Coding Scheme for Distributed Computation of Approximated Matrix Multiplication}
\author{Tayyebeh~Jahani-Nezhad, Mohammad~Ali~Maddah-Ali 
\thanks{
Tayyebeh Jahani-Nezhad is with  the  Department of Electrical Engineering, Sharif University of Technology, Tehran, Iran. (email:tayyebeh.jahaninezhad@ee.sharif.edu).
Mohammad Ali Maddah-Ali is with    the  Department of Electrical Engineering, Sharif University of Technology, Tehran, Iran. (email: maddah\_ali@sharif.ir).}
	\thanks{
	This paper has been partially presented  at  IEEE ISIT 2019 \cite{CodedSketch}.
}
 }
 \renewcommand\footnotemark{}
\maketitle

\begin{abstract} 
	In this paper, we propose \emph{CodedSketch},  as a distributed straggler-resistant scheme to compute an approximation of the multiplication of two massive matrices.  The objective is to reduce the \emph{recovery threshold}, defined as the total number of worker nodes that the master node needs to wait for to be able to recover the final result.  To exploit the fact that only an approximated result is required, in reducing the recovery threshold,  some sorts of pre-compression are required. However, compression inherently involves some randomness that would lose the structure of the matrices. On the other hand, considering the structure of the matrices is crucial to reduce the recovery threshold. In CodedSketch, we use count--sketch, as a hash-based compression scheme,  on the rows of the first and columns of the second matrix, and a structured polynomial code on the columns of the first and rows of the second matrix. This arrangement allows us to exploit the gain of both in reducing the recovery threshold. To increase the accuracy of computation, multiple independent count--sketches are needed. This independency allows us to theoretically characterize the accuracy of the result and establish the recovery threshold achieved by the proposed scheme.  To guarantee the independency of resulting count--sketches in the output, while keeping its cost on the recovery threshold minimum, we use another layer of structured codes.  The proposed scheme provides an upper-bound on the recovery threshold as a function of the required accuracy of computation and the probability that the required accuracy can be violated. In addition, it provides an upper-bound on the recovery threshold for the case that the result of the multiplication is sparse, and the exact result is required.  
\end{abstract}


\section{Introduction}
Linear operations, often represented by matrix multiplication, are the key techniques used in many applications such as optimization and machine learning.  
 Many such applications require processing large-scale matrices. For example,  in the deep neural networks, the convolution layers, which are  operations based on matrix multiplication, account for a large fraction of computational time~\cite{conv}. To increase the accuracy of the learning, we increase the size of the model,  by using more layers and more neurons in each layer. This would increase the computation complexity and the overall training time. This heavy computation cannot be completed over a single machine.  An inevitable solution to overcome this challenge is to distribute computation  over several machines\cite{map,spark}.  In particular, consider a master-worker setting for distributed computing, in which the master node has access to the information of two matrices. The master then partitions the task of matrix multiplications into some smaller sub-tasks and assigns each sub-task to one of the worker nodes to be executed.  In this setting, the execution time is dominated by the speed of the slowest worker nodes, or the \emph{stragglers}. This is one of the main challenges in distributed computing~\cite{tail}. 

Conventionally, the effect of stragglers has been mitigated by using redundancy in computing. This means that every task is executed over more than one machine, and the results are fetched  from the fastest ones.  More recently, it is shown that coding can be more effective in coping with stragglers in terms of minimizing the \emph{recovery threshold}, defined as the total number of worker nodes that the master node needs to wait for to be able to recover the final result~\cite{speed,high,Polynomial,entangle, opt-recovery, short, Multi-Party, Multi-Party2, sketch, sparse, lagrange}.  In~\cite{speed,high}, it is suggested that in matrix multiplication, one or both of the matrices are coded separately using maximum distance separable (MDS) codes. In \cite{Polynomial},  \emph{polynomial codes} have been proposed to code each matrix, such that the result of the multiplications across worker nodes become also MDS coded.  
The \emph{short-dot} technique has been presented in \cite{short},  where  coded redundancy is added to the computation, and the effort is to keep the coded matrices sparse and thus reduce the load of the computation. In \cite{entangle}, an extension of the polynomial codes, known as \emph{entangled polynomial codes}, has been proposed that admits flexible partitioning of each matrix and minimizes the number of unwanted computations. The general version of entangled polynomial codes is proposed in~\cite{opt-recovery}, this means a coding scheme is introduced to multiply more than two matrices.
In~\cite{sparse}, coded sparse matrix multiplication is proposed for the case where the result of the multiplication is sparse. In~\cite{Multi-Party, Multi-Party2}, coded schemes have been used to develop multi-party computation scheme to calculate arbitrary polynomials of massive matrices, while preserving privacy of the data. In~\cite{lagrange}, a universal  coding technique, \emph{Lagrange Code}, is developed to code across several parallel computations of an arbitrary polynomial function, without communication across worker nodes. In~\cite{Grad_Dimakis},
a coded computing framework is proposed to mitigate the effect of stragglers in the distributed gradient descent algorithm.  Since in SGD, the algorithm works even with the approximate result, in~\cite{grad_app_dimakis} an approximate variant of the gradient coding in a straggler-resistant distributed setting is introduced using expander graphs. In~\cite{grad_app},  a fundamental trade-off among the computation load, error of computed gradient, and the number of stragglers is proposed. Similarly,  there are many applications which process large-scale matrices, and computing the exact result of matrix multiplication is not required. In those cases, computing approximated matrix multiplication may significantly decrease the required number of worker nodes and computation load. 
To derive an approximated result of the matrix multiplication over a single server,  linear sketching as a randomized method can be applied  \cite{woodruff2014sketching,mahoney2011randomized,clarkson2009numerical}. Linear sketching is an approach to reduce the dimension of the input matrices by multiplying them by a random matrix with certain properties. Also, in~\cite{Compressed}, it is proposed to  use the count-sketch  method,  concatenated with fast Fourier transform to reduce multiplication complexity. Count-Sketch is a hashing-based approach of developing linear sketching to reduce the size of the input vector such that it  can be recovered with a certain accuracy. Count-sketch method can be particularly efficient in massive computing, machine learning and statistics \cite{wang2017sketched,liu2020learned}.  To review count-sketch  method, see~\cite{findfrequent2,findfrequent}, or Subsection~II-A on preliminaries in this paper.
  In~\cite{sketch},  the \emph{OverSketch} method has been proposed where some extra count-sketches are used, as a redundancy,  to mitigate  stragglers in a distributed setting (see Remark~\ref{remark5} for further discussion).

%

\begin{figure}
		\centerline{\includegraphics[scale=1,draft=false]{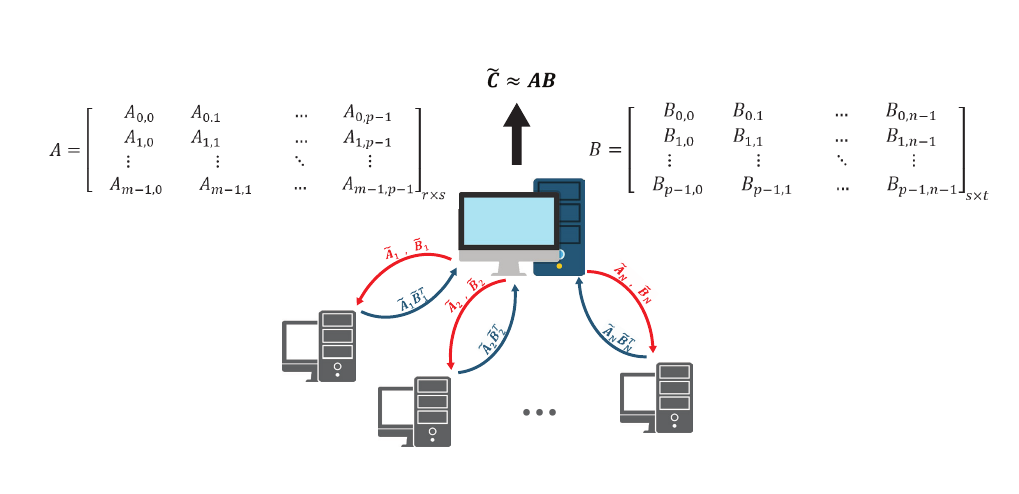}}
		\caption{Framework of distributed computation of approximated matrix multiplications.
		}
\label{fig44}
\end{figure}

In this paper, we propose a distributed straggler--resistant computation scheme to achieve an approximation of the multiplication of two large matrices $\mathbf{A}\in \mathbb{R}^{r\times s}$ and $\mathbf{B}\in \mathbb{R}^{s \times t}$, where $r, s, t \in \mathbb{N}$ (see Fig.~\ref{fig44}).  The motivation is that in some optimization and machine learning algorithms which are based on stochastic gradient descent such as the learning phase of deep learning, the scheme is less sensitive to the accuracy of the result as long as it is an unbiased estimation of it. To exploit the fact that an approximated result is required to reduce the recovery threshold,  we need to use some sorts of pre-compression. However, compression inherently involves some randomness that would lose the structure of the matrices. On the other hand, considering the structure of the matrices is crucial to reduce the recovery threshold. In this paper, we use the ideas of entangled polynomial codes and count-sketch to form a right combination of randomness and structured codes.	To be specific, we use entangled polynomial codes on the columns of $\mathbf{A}$ and the rows of $\mathbf{B}$. Then,
we use count--sketch compression on the rows of $\mathbf{A}$ and the columns of $\mathbf{B}$. This arrangement allows us to enjoy the benefits of both in reducing the recovery threshold.  To improve the overall accuracy, we need to use multiple count--sketches. Another layer of structured codes, i.e., improved entangled polynomial codes, allows us to keep multiple count--sketches independent. This independency is used to prove theoretical guarantee on the performance of the final result and establish an achievable  recovery threshold.

\textit{Notation}: For $n_1, n_2 \in \mathbb{Z}$, the notation $[n_1 : n_2]$  represents  the set $\{n_1, n_1 + 1, ..., n_2\}$. The cardinality of a set $\mathcal{S}$ is denoted by $|\mathcal{S}|$. In addition, $\mathbf{E}[X]$ refers to the expected value of $X$. The $i$th element of a vector $\mathbf{v}$, is denoted by ${v}_i$  and  the $(i,j)$-th entry of a matrix $\mathbf{A}$ is denoted by $[\mathbf{A}]_{i,j}$. Also, $\mathbf{A}_{k,\ell}$ is block $(k,\ell)$ of block matrix $\mathbf{A}$, and $[\mathbf{A}_{k,\ell}]_{i,j}$ is entry $(i,j)$ in the block $(k,\ell)$ of block matrix $\mathbf{A}$.  $||\mathbf{A}||_F$ is the Frobenius norm of matrix $\mathbf{A}$.

\section{Problem Formulation, Notation and Preliminaries}\label{ProbFor}
\subsection{Preliminaries}\label{prelim}
In this subsection, we review some preliminaries. 
\begin{definition} [$k$-wise Independent Random Variables \cite{Probability}]
	A set of discrete random variables $X_1, X_2,..., X_n$ is $k$-wise independent if for any subset $\mathcal{S}$ of these random variables, $|\mathcal{S}|\le k$,  and for any values $x_i$, $i \in [1:|\mathcal{S}|]$  we have 
	\begin{equation}
	\mathbf{P}(\bigcap\limits_{i = 1}^{|\mathcal{S}|} {{X_i} = {x_i}} ) = \prod\limits_{i = 1}^{|\mathcal{S}|} {\mathbf{P}({X_i} = {x_i})}.
	\end{equation}
\end{definition}
\begin{definition} [Hash Function \cite{Probability}]
A hash function is a function that maps a universe $\mathcal{U}$ into a special range $[0:b-1]$ for some $b\in \mathbb{N}$, i.e., $h:\mathcal{U}\to [0:b-1]$. In other words, a hash function operates as a method in which items from the universe are placed into $b$ bins.
\end{definition}
The most simple family of hash functions  is \emph{completely random} hash functions which can be analyzed using a model known as \emph{balls and bins} model. The hash values $h(x_1), h(x_2),...,h(x_n)$ are considered independent and uniform over the range of the hash function for any collection of data $x_1, x_2,...,x_n$.  Since this class of hash functions is expensive in terms of computing and storage cost, it is not very useful in practice.
Another family of hash functions is known as \emph{Universal family}, which satisfies some provable performance guarantees \cite{Probability}.
\begin{definition}[$k$-universal Hash Family \cite{Probability}]
The hash family $\mathcal{H}$ is defined as a $k$-universal family if for any hash function  $h:\mathcal{U}\to [0:b-1]$ with $|\mathcal{U}|\ge b$,  chosen uniformly at random from $\mathcal{H}$, and for any collection of $x_1, x_2,...,x_k$, we have
\begin{equation}
\mathbf{P}(h(x_1)=h(x_2)=...=h(x_k))\le \frac{1}{b^{k-1}}.
\end{equation}
In addition $\mathcal{H}$ is defined as a \emph{strongly} $k$-universal family if for any values $y_1,...,y_k \in [0:b-1]$ we have
\begin{equation}
\mathbf{P}((h(x_1)=y_1)\cap (h(x_2)=y_2)\cap ...\cap (h(x_k)=y_k))= \frac{1}{b^{k}}.
\end{equation}
This definition implies that  for any fix $x\in \mathcal{U}$,  $h(x)$ is uniformly distributed over $[0:b-1]$. Also, if $h$ is chosen uniformly at random from $\mathcal{H}$, for any distinct $x_1, x_2,...,x_k$, the values of $h(x_1),h(x_2),...,h(x_k)$ are independent.
\end{definition}
\begin{remark}
	The values $h(x_1), h(x_2),...,h(x_k)$ are $k$-wise independent if the hash function $h$ is chosen from a strongly $k$-universal family.
\end{remark}
\begin{definition} [Count-Sketch \cite{findfrequent2, findfrequent}]
The count-sketch is a method for representing a compact form of data which maps an $n$-dimensional vector $\mathbf{a}$ to a $b$-dimensional vector $\mathbf{c}$ where $n\gg b$. The  count-sketch of $\mathbf{a}$ can be defined using an iterative process, initialized  with the vector $\mathbf{c}$ to be entirely zero,  as follows: in the $i$th iteration ($i=[0:n-1]$) we have 
\begin{equation}
{c}_{h(i)}={c}_{h(i)}+s(i)a_i,
\end{equation}
where $s:[0:n-1]\to\{-1,1\}$ is a sign function, $a_i$ is $i$th entry of $\mathbf{a}$, and $h:[0:n-1] \to [0:b-1]$ is a hash function that $h(i)$ computes the hash of $i$. 
\end{definition}

One count-sketch of $\mathbf{a}$ is created by using one hash function and one sign function.  To improve accuracy in this approximation, more than one hash function can be used (say $d \in \mathbb{N}$ hash functions).  To create $d$ count-sketches, we need  $d$ pairwise independent sign functions and $d$ 2-wise independent hash functions.  The output of these $d$ count-sketches is a $d\times b$ matrix.  $a_i$ as the entry $i$ of $\mathbf{a}$, for each $i \in [0:n-1]$, can be approximately recovered by taking  the median of the values of some entries of this matrix. For details, see  Algorithm \ref{alg}.
\begin{theorem}[\cite{findfrequent2, lecture}]\label{Th1}
	If $d\ge \log\frac{1}{\delta}$ and $b\ge\frac{3}{\epsilon^2}$ then  for entry $a_i$ of the input vector $\mathbf{a}$, $i\in [0:n-1]$, we have 
	\begin{align}
	\mathbf{E}[c_{t,i}]& =a_i, \\
	\mathbf{Var}[c_{t,i}] & \le \frac{{\left\| a \right\|_2^2}}{b},
	\end{align}
	and
	\begin{equation}
\mathbf{P}\bigg[|c_{t,i}-a_i|\ge \epsilon{\left\| \mathbf{a} \right\|_2}\bigg]\le \frac{1}{3}
	\label{eq12}
	\end{equation}
	where $c_{t,i}=s_t(i)[\mathbf{C}]_{t,h_t(i)}$ is an estimation of $a_i$. 
\end{theorem}
	\begin{corollary}[\cite{findfrequent2, lecture}]\label{cor1}
	If there are $d$ estimations like $c_{t,i}$, then  
	\begin{equation}
			\mathbf{P}\bigg[|\textnormal{median}\{c_{1,i}, c_{2,i},...,c_{d,i}\}-a_i|\ge \epsilon{{{\left\| \mathbf{a} \right\|}_2}}\bigg]\le \delta.
			\label{eq13}
	\end{equation}
	\end{corollary}
Corollary~\ref{cor1} shows that by choosing $d=\mathcal{O}(\log n)$ with probability at least $1-\frac{1}{\text{poly}(n)}$, we have $|\tilde{a}_i-a_i|\le \epsilon {{{\left\| \mathbf{a} \right\|}_2}}$, where $\tilde{a}_i=\text{median}\{c_{1,i}, c_{2,i},...,c_{d,i}\}$, for $i\in [0:n-1]$.
\begin{remark}
	Suppose $n$-dimensional vector $\mathbf{a}$ has only $\kappa$ nonzero entries. Then, from~Corollary~\ref{cor1}, one can see that an approximated version of $\mathbf{a}$ with accuracy $\epsilon=\mathcal{O}(\frac{1}{\sqrt \kappa})$ can be achieved by $d=\mathcal{O}(\log n)$ count-sketches. 
\end{remark}
\begin{theorem}[\cite{Improved,stanford,lecture}]\label{th2}
	For an $n$-dimensional vector $\mathbf{a}$ and  $\kappa \in [0:n]$, if we choose the length of the count-sketches as $b\ge \frac{3\kappa}{\epsilon^2}$ and the number of count-sketches as $d$ then we have
	\begin{equation}
	\mathbf{P}\bigg[|\tilde{a}_i-{a}_i|\ge \frac{\epsilon}{\sqrt{\kappa}}\mathrm{err}_2^{(\kappa)}(\mathbf{a})\bigg]\le 2e^{-\Omega(d\epsilon^2)}, 
	\end{equation}
	where  $\mathrm{err}_2^{(\kappa)}(\mathbf{a})={(\sum_{i \notin \mathcal{S}}{|a_i|^2})}^\frac{1}{2}$ and $\mathcal{S}$ is the set of  indices of the $\kappa$ largest (in absolute value) entries of $\mathbf{a}$.
\end{theorem}
\begin{remark}\label{remark3}
	According to Theorem \ref{th2}, if the $n$-dimensional vector $\mathbf{a}$ has only $\kappa$ nonzero entries and $d=\mathcal{O}(\log n)$, then the output of the count-sketch method computes $\mathbf{a}$ exactly with probability $1-\frac{1}{n^{\Theta(1)}}$. 
\end{remark}

\begin{definition} \label{def5}
	The count-sketch of an $n$-dimensional vector $\mathbf{a}$ can be represented by a polynomial, named \emph{sketch polynomial}, as 
	\begin{equation}
	{p_\mathbf{a}}(\alpha) = \sum\limits_{i = 0}^{n-1} {s(i){a_i}{\alpha^{h(i)}}},
	\end{equation}
	where $h:[0:n-1]\to[0:b-1]$ is the 2-wise independent hash function and $s:[0:n-1]\to \{-1,1\}$ is the 2-wise independent sign function, used to develop the count-sketch. 
\end{definition}

Let us assume that we use $h_1$ and $s_1$ to count-sketch vector $\mathbf{a}$ and $h_2$ and $s_2$ to count-sketch vector $\mathbf{b}$, respectively, 
represented by sketch polynomials 
\begin{align}
{p_\mathbf{a}}(\alpha) & = \sum\limits_{i = 0}^{n-1} {s_1(i){a_i}{\alpha^{h_1(i)}}},\\
{p_\mathbf{b}}(\alpha)  & = \sum\limits_{i = 0}^{n-1} {s_2(i){b_i}{\alpha^{h_2(i)}}},  
\end{align}
then ${p_{\mathbf{a}\mathbf{b}^T}}(\alpha) \triangleq {p_\mathbf{a}}(\alpha) {p_\mathbf{b}}(\alpha)$ represents a sketch polynomial for matrix $\mathbf{a}\mathbf{b}^T$, as 
%
\begin{equation}
{p_{\mathbf{a}\mathbf{b}^T}}(\alpha) = \sum\limits_{i,j} {s(i,j){{[\mathbf{a}\mathbf{b}^T]}_{ij}}{\alpha^{h(i,j)}}}.
\label{eq2}
\end{equation}
where
\begin{equation}
s(i,j)=s_1(i)s_2(j). 
\end{equation}
Also, the hash function on pairs $(i,j)$ is
\begin{equation}
h(i,j)=h_1(i)+h_2(j). 
\end{equation}
\begin{remark}
	Recall that $h_1(i)$ and $h_2(j)$ are two independent random variables with some distributions. Thus the distribution of $h(i,j)$ over its output range is the convolution of the distributions of $h_1(i)$ and $h_2(j)$.

%
\end{remark}	
\begin{algorithm}
	\caption{The count-sketch algorithm \cite{findfrequent2, findfrequent}}
	\label{alg} 
		\begin{algorithmic}[1]
		\Function{count-sketch($\mathbf{a}$)}{}
		\State $\mathbf{a}$: an $n$-dimensional input vector
		\State $\mathbf{C}$: a $d\times b$ matrix, initialized by $\mathbf{0}$
		\State choose $h_1, h_2, ..., h_d:[0:n-1]\to[0:b-1]$ from the family of 2-wise independent  of hash functions
		\State choose $s_1, s_2, ..., s_d:[0:n-1]\to\{-1,1\}$ from the family of 2-wise independent functions 
		\For{$t=1$ to $d$}
		\For{$i = 0$ to $n-1$}
		\State $[\mathbf{C}]_{t,h_t(i)} \leftarrow [\mathbf{C}]_{t,h_t(i)}+s_t(i)a_i$
		\EndFor	
		\EndFor
		\EndFunction
		\Function{recovering $\tilde{\mathbf{a}}$ from $\mathbf{C},h,s$}{}
		\For{$j=0$ to $n-1$}
		\For{$t=1$ to $d$}
		\State $c_{t,j}=s_t(j)[\mathbf{C}]_{t,h_t(j)}$
		\EndFor
		\State $\tilde{a}_j=\mathrm{median}\{{c_{t,j}}\}^{d}_{t=1}$
		\EndFor
		\EndFunction
	\end{algorithmic}
\end{algorithm}
\subsection{Problem Formulation}
Consider a distributed system including a master and $N$ worker nodes, where the master node is connected to each worker node. Assume that the master node wants to compute $\tilde{\mathbf{C}}$ which is an approximation of the multiplication of two matrices $\mathbf{C}=\mathbf{A}\mathbf{B}$, where $\mathbf{A}\in \mathbb{R}^{r\times s}$, $\mathbf{B}\in \mathbb{R}^{s \times t}$, for some positive integers $ m, n$ and $p$, are partitioned into $m\times p$ and  $p\times n$ sub-matrices of equal size, respectively. Our goal is to compute $\tilde{\mathbf{C}}$ subject to the following conditions
\begin{enumerate}
	\item 
	Unbiasedness:
\begin{equation}\label{unbias}
\mathbf{{E}}[\tilde{\mathbf{C}}]=\mathbf{C}.
\end{equation}
\item
$(\epsilon, \delta)$-accuracy:\\
	We say $\tilde{\mathbf{C}}$ is an $(\epsilon, \delta)$-accurate approximation of matrix $\mathbf{C}$ if
	 \begin{align}\label{epsilon-delta}
	\mathbf{P}\bigg[|[\mathbf{C}]_{i,j}-[\tilde{\mathbf{C}}]_{i,j}|\ge{\epsilon}||\mathbf{C}||_{F,(i',j')} \bigg]\le\delta\hspace{1cm} \forall i\in[0:r-1],\forall j\in[0:t-1],
	\end{align}
	where $i'=i \mod \frac{r}{m}$, $j'=j \mod \frac{t}{n}$, and $[\mathbf{C}]_{i,j}$ is the $(i,j)$th entry of $\mathbf{C}$. Also, we define
	\begin{align}
	||\mathbf{C}||_{F,(i',j')}\triangleq(\sum_{k=0}^{m-1}\sum_{\ell=0}^{n-1}{[\mathbf{C}_{k,\ell}]_{i',j'}^2})^\frac{1}{2}.
		\end{align}
\end{enumerate}
\begin{remark}
	In \eqref{unbias} and \eqref{epsilon-delta}, the randomness is over the count-sketches which are used in the proposed scheme.
	\end{remark}
 Suppose $\mathbf{A}$ and $\mathbf{B}$ are partitioned into the sub-matrices of equal size as
\begin{equation}
	\mathbf{A} = {\left[ {\begin{array}{*{20}{c}}
			{{\mathbf{A}_{0,0}}}&{{\mathbf{A}_{0,1}}}&{...}&{{\mathbf{A}_{0,p - 1}}}\\
			{{\mathbf{A}_{1,0}}}&{{\mathbf{A}_{1,1}}}&{...}&{{\mathbf{A}_{1,p - 1}}}\\
			\vdots & \vdots & \ddots & \vdots \\
			{{\mathbf{A}_{m - 1,0}}}&{{\mathbf{A}_{m - 1,1}}}& \cdots &{{\mathbf{A}_{m - 1,p - 1}}}
			\end{array}} \right]_{r \times s}},
	\label{eq7}
\end{equation}
\begin{equation}
\label{eq8}
\mathbf{B} = {\left[ {\begin{array}{*{20}{c}}
		{{\mathbf{B}_{0,0}}}&{{\mathbf{B}_{0,1}}}&{...}&{{\mathbf{B}_{0,n - 1}}}\\
		{{\mathbf{B}_{1,0}}}&{{\mathbf{B}_{1,1}}}&{...}&{{\mathbf{B}_{1,n - 1}}}\\
		\vdots & \vdots & \ddots & \vdots \\
		{{\mathbf{B}_{p - 1,0}}}&{{\mathbf{B}_{p - 1,1}}}& \cdots &{{\mathbf{B}_{p - 1,n - 1}}}
		\end{array}} \right]_{s \times t}},
\end{equation}
where $m, n$ and $p$ are positive integers. One of the  constraints in this system is the limited storage of each worker. We assume that the size of the storage at each node is equal to $\frac{1}{pm}$ fraction of $\mathbf{A}$ plus $\frac{1}{pn}$ fraction of $\mathbf{B}$. Assume $\mathcal{E}_i:\mathbb{R}^{r\times s}\to \mathbb{R}^{\frac{r}{m}\times \frac{s}{p}}$  and $\tilde{\mathcal{E}}_i:\mathbb{R}^{s\times t}\to \mathbb{R}^{\frac{s}{p}\times \frac{t}{n}}$ are the encoding functions that are used by the master node to compute $\tilde{\mathbf{A}}_i$ and $\tilde{\mathbf{B}}_i$ for $i$th worker node, $i=1,...,N$, respectively. In other words, the master node encodes the input matrices and sends two coded matrices to $i$th worker node as follows:
\begin{align}
\tilde{\mathbf{A}}_i={\mathcal{E}}_i({\mathbf{A}}),\\
\tilde{\mathbf{B}}_i=\tilde{\mathcal{E}}_i({\mathbf{B}}),
\end{align}
where  $\tilde{\mathbf{A}}_i\in \mathbb{R}^{\frac{r}{m}\times \frac{s}{p}}$ and $\tilde{\mathbf{B}}_i\in \mathbb{R}^{\frac{s}{p}\times \frac{t}{n}} $. 
Each worker node computes $\tilde{\mathbf{C}}_i\triangleq\tilde{\mathbf{A}}_i\tilde{\mathbf{B}}_i$ and sends back the result to the master node. The problem formulation so far limits the communication and computation load per node. After receiving the results from a subset $\mathcal{K} \subset[N]$ of worker nodes, the master node can recover $\tilde{\mathbf{C}}$, the approximation of the original result. Let $\mathcal{D}(.)$ be a reconstruction function which operates on a subset $\mathcal{K}$ of the workers' results and calculates $\tilde{\mathbf{C}}$ as follows
\begin{equation}
\tilde{\mathbf{C}}=\mathcal{D}(\{\tilde{\mathbf{C}}_i\}_{i\in\mathcal{K}}).
\end{equation} 
\begin{definition}
	The recovery threshold of a distributed system with $N$ worker nodes and one master node is the minimum number of workers that the master node needs to wait for such that it is guaranteed the master node can complete the computation, subject to space constraint at each worker node and unbiasedness and $(\epsilon, \delta)$-accuracy conditions. In this paper, the \emph{optimum} recovery threshold is denoted by $N^{*}(\epsilon,\delta,p,m,n)$.
\end{definition}
\section{Main results}
The following theorems state the main results of the paper.
	\begin{theorem}\label{th3}
	 For $(\epsilon, \delta)$-accurate approximation of the multiplication of two matrices, we have 
	\begin{equation}
	N^{*}(\epsilon,\delta,p,m,n)  \leq
	 \min\left\{ {(2p\left\lceil {\frac{3}{{{\epsilon ^2}}}} \right\rceil  - 1)(2\left\lceil {\log \frac{1}{\delta }} \right\rceil  - 1) - 1}, pmn+p-1 \right\}.
	\end{equation}
		\end{theorem}
The proof of Theorem~\ref{th3} is detailed in Subsection~\ref{proof4}.

\begin{remark}
	To prove Theorem \ref{th3}, we propose a coding scheme that benefits from three different features: 
	\begin{enumerate}
		\item It exploits the fact that only an approximation result is needed, by using some count-sketches to pre-compress the input matrices which reduces the recovery threshold. 
		
		\item It relies on the structure of the matrix multiplication to add coded redundancy in the computation and mitigate the effect of stragglers. On the other hand, this coding can force the system to execute some unwanted calculations. The coding is designed such that the total number of unwanted calculations is reduced while the coding is maximum distance separable (MDS). 
		
		\item  The proposed scheme, in the end, creates some independent count-sketches of $\mathbf{A}\mathbf{B}$, from which we calculate the final result.   As a side product,  some dependent count-sketches  are also created. To minimize the recovery threshold,  we need to minimize the number of these side products.  Another layer of structured code is used to reduce the number of these side products and thus reduce the recovery threshold.
		
	\end{enumerate}
	
	We note that between opportunities one and two, the first one injects some randomness to the input matrices, while the second one relies on their structure. To achieve both at the same time, we use count-sketches on the rows of the first and columns of the second matrix, and use a structured code on the columns of the first and the rows of the second matrix. 
	
	The code structure is motivated by count-sketch~\cite{findfrequent2,findfrequent}, polynomial codes~\cite{Polynomial}, entangled polynomial codes~\cite{entangle}, MatDot codes~\cite{opt-recovery}, and improved entangled polynomial codes.	
\end{remark}

\begin{remark}
	The proposed scheme achieves  the first term in Theorem \ref{th3}.  The second term is achieved by the entangled polynomial code~\cite{entangle} to calculate the exact result. 
	\end{remark}

\begin{remark}
  The proposed scheme computes the approximated matrix multiplication in a straggler-resistant distributed system using one-shot coded computation, this means in this scheme one encoding, and one decoding step is used which will be explained below. In this scheme, no communication between worker nodes is needed, and the master node only sends two coded matrices to each worker node, and each worker node returns its computation result to the master node without any additional communication between worker nodes.
	\end{remark}
\begin{remark}
In the proposed scheme, the amount of computation needed at the master node to find the approximate multiplication result of two $n\times n$ matrices scales $\mathcal{O}(n^2)$. Also, the problem formulation restricts the computational complexity of each worker node which means a fixed computation load is specified to each worker node, and it only receives $\frac{1}{pm}$ of the matrix $\mathbf{A}$ and $\frac{1}{pn}$ of the matrix $\mathbf{B}$ and returns the multiplication of these two matrices to the master node.
The computational complexity of the proposed scheme in the master node and each worker node is detailed in Subsection \ref{COMP}.
\end{remark}
\begin{theorem}\label{th4}
	If matrix $\mathbf{C}=\mathbf{A}\mathbf{B}$ is $\kappa$-sparse, i.e., for all $i\in[0:\frac{r}{m}-1]$ and $j\in[0:\frac{t}{n}-1]$, there exist at most $\kappa$ non-zero elements among the $(i,j)$th entries of all $mn$ blocks of  $\mathbf{C}$, then the proposed method computes $\mathbf{C}$ exactly with probability $1\hspace{-0.5mm}-\hspace{-0.5mm}\delta\hspace{-0.5mm}=\hspace{-0.5mm}1\hspace{-0.5mm}-\hspace{-0.5mm}\frac{1}{(mn)^{\Theta(1)}}$ and with the recovery threshold of
	\begin{equation}
	\min\left\{(2p\left\lceil {\frac{3\kappa}{{{\epsilon^2}}}} \right\rceil - 1)(2\left\lceil {\log {mn} } \right\rceil  - 1) - 1, pmn+p-1 \right\}.
	\end{equation}	
\end{theorem}

The proof of  Theorem~\ref{th4} is detailed in Subsection~\ref{proof5}.  

\section{The Proposed Scheme}\label{propose}
In the following, the goal is to compute the approximation of the matrix multiplication over a distributed system using CodedSketch scheme.
\subsection{Motivating Example}

We first demonstrate the main idea of our scheme through a simple example.  We then generalize this approach in the next section. 
Consider a distributed system with one master node and $N$ worker nodes which aim to collaboratively compute $\tilde{\mathbf{C}}$ as an approximation of $\mathbf{C}=\mathbf{A}\mathbf{B}$ where $\mathbf{A}$ and $\mathbf{B}$ are two matrices partitioned as follows
\begin{equation}
\mathbf{A} = \left[ {\begin{array}{*{20}{c}}
	{{\mathbf{A}_{0,0}}}&{{\mathbf{A}_{0,1}}}&{{\mathbf{A}_{0,2}}}&{{\mathbf{A}_{0,3}}}\\
	{{\mathbf{A}_{1,0}}}&{{\mathbf{A}_{1,1}}}&{{\mathbf{A}_{1,2}}}&{{\mathbf{A}_{1,3}}}\\
	{{\mathbf{A}_{2,0}}}&{{\mathbf{A}_{2,1}}}&{{\mathbf{A}_{2,2}}}&{{\mathbf{A}_{2,3}}}\\
	{{\mathbf{A}_{3,0}}}&{{\mathbf{A}_{3,1}}}&{{\mathbf{A}_{3,2}}}&{{\mathbf{A}_{3,3}}}
	\end{array}} \right]_{r\times s},
\end{equation}
\begin{equation}
\mathbf{B} = \left[ {\begin{array}{*{20}{c}}
	{{\mathbf{B}_{0,0}}}&{{\mathbf{B}_{0,1}}}&{{\mathbf{B}_{0,2}}}&{{\mathbf{B}_{0,3}}}\\
	{{\mathbf{B}_{1,0}}}&{{\mathbf{B}_{1,1}}}&{{\mathbf{B}_{1,2}}}&{{\mathbf{B}_{1,3}}}\\
	{{\mathbf{B}_{2,0}}}&{{\mathbf{B}_{2,1}}}&{{\mathbf{B}_{2,2}}}&{{\mathbf{B}_{2,3}}}\\
	{{\mathbf{B}_{3,0}}}&{{\mathbf{B}_{3,1}}}&{{\mathbf{B}_{3,2}}}&{{\mathbf{B}_{3,3}}}
	\end{array}} \right]_{s\times t}.
\end{equation}
The result of the multiplication can be computed using summation of four outer products as $\mathbf{C}=\sum_{k=0}^{3}{\mathbf{A}_k\mathbf{B}^T_k}$, where $\mathbf{A}_k$ and $\mathbf{B}_k^T$ are the $k$th column of $\mathbf{A}$ and $k$th row of $\mathbf{B}$ respectively. The proposed scheme is based on the following steps.\\
\begin{itemize}
	\item
	\textit{\textbf{Step 1}}.
The master node forms the following polynomial matrices based on column-blocks of $\mathbf{A}$ and row-blocks of $\mathbf{B}$ as follows
\begin{equation}\label{23}
\begin{aligned}
\hat{\mathbf{A}}(x) =\left[ {\begin{array}{*{20}{c}}
		{{{\hat{\mathbf{A}}}_{0,0}}(x)}\\
		{{{\hat{\mathbf{A}}}_{1,0}}(x)}\\
		{{{\hat {\mathbf{A}}}_{2,0}}(x)}\\
		{{{\hat {\mathbf{A}}}_{3,0}}(x)}
		\end{array}} \right] &\buildrel \Delta \over =  \left[ {\begin{array}{*{20}{c}}
		{{\mathbf{A}_{0,0}}}\\
		{{\mathbf{A}_{1,0}}}\\
		{{\mathbf{A}_{2,0}}}\\
		{{\mathbf{A}_{3,0}}}
		\end{array}} \right] + x\left[ {\begin{array}{*{20}{c}}
		{{\mathbf{A}_{0,1}}}\\
		{{\mathbf{A}_{1,1}}}\\
		{{\mathbf{A}_{2,1}}}\\
		{{\mathbf{A}_{3,1}}}
		\end{array}} \right] + {x^2}\left[ {\begin{array}{*{20}{c}}
		{{\mathbf{A}_{0,2}}}\\
		{{\mathbf{A}_{1,2}}}\\
		{{\mathbf{A}_{2,2}}}\\
		{{\mathbf{A}_{3,2}}}
		\end{array}} \right] + {x^3}\left[ {\begin{array}{*{20}{c}}
		{{\mathbf{A}_{0,3}}}\\
		{{\mathbf{A}_{1,3}}}\\
		{{\mathbf{A}_{2,3}}}\\
		{{\mathbf{A}_{3,3}}}
		\end{array}} \right]\\
	&=\left[ {\begin{array}{*{20}{c}}
		{{\mathbf{A}_{0,0}} + x{\mathbf{A}_{0,1}} + {x^2}{\mathbf{A}_{0,2}}+{x^3}{\mathbf{A}_{0,3}}}\\
		{{\mathbf{A}_{1,0}} + x{\mathbf{A}_{1,1}} + {x^2}{\mathbf{A}_{1,2}}+{x^3}{\mathbf{A}_{1,3}}}\\
		{{\mathbf{A}_{2,0}} + x{\mathbf{A}_{2,1}} + {x^2}{\mathbf{A}_{2,2}}+{x^3}{\mathbf{A}_{2,3}}}\\
		{{\mathbf{A}_{3,0}} + x{\mathbf{A}_{3,1}} + {x^2}{\mathbf{A}_{3,2}}+{x^3}{\mathbf{A}_{3,3}}}
		\end{array}} \right],
\end{aligned}
\end{equation}
and
\begin{equation}\label{24}
\begin{aligned}
\hat{\mathbf{B}}(x)^T={\left[ {\begin{array}{*{20}{c}}
		{\hat {\mathbf{B}}_{0,0}(x)}\\
		{\hat {\mathbf{B}}_{0,1}(x)}\\
		{\hat {\mathbf{B}}_{0,2}(x)}\\
		{\hat {\mathbf{B}}_{0,3}(x)}
		\end{array}} \right]^T}\hspace{-3mm} &\buildrel \Delta \over ={x^3}{\left[ {\begin{array}{*{20}{c}}
		{\mathbf{B}_{0,0}^T}\\
		{\mathbf{B}_{0,1}^T}\\
		{\mathbf{B}_{0,2}^T}\\
		{\mathbf{B}_{0,3}^T}
		\end{array}} \right]^T}\hspace{-2mm}+{x^2}{\left[ {\begin{array}{*{20}{c}}
		{\mathbf{B}_{1,0}^T}\\
		{\mathbf{B}_{1,1}^T}\\
		{\mathbf{B}_{1,2}^T}\\
		{\mathbf{B}_{1,3}^T}
		\end{array}} \right]^T}\hspace{-2mm}+x{\left[ {\begin{array}{*{20}{c}}
		{\mathbf{B}_{2,0}^T}\\
		{\mathbf{B}_{2,1}^T}\\
		{\mathbf{B}_{2,2}^T}\\
		{\mathbf{B}_{2,3}^T}
		\end{array}} \right]^T}\hspace{-2mm}+{\left[ {\begin{array}{*{20}{c}}
		{\mathbf{B}_{3,0}^T}\\
		{\mathbf{B}_{3,1}^T}\\
		{\mathbf{B}_{3,2}^T}\\
		{\mathbf{B}_{3,3}^T}
		\end{array}} \right]^T}\hspace{-2mm}\\
&={\left[ {\begin{array}{*{20}{c}}
		{{x^3}\mathbf{B}_{0,0}^T + {x^2}\mathbf{B}_{1,0}^T + x\mathbf{B}_{2,0}^T+\mathbf{B}_{3,0}^T}\\
		{{x^3}\mathbf{B}_{0,1}^T + {x^2}\mathbf{B}_{1,1}^T + x\mathbf{B}_{2,1}^T+\mathbf{B}_{3,1}^T}\\
		{{x^3}\mathbf{B}_{0,2}^T + {x^2}\mathbf{B}_{1,2}^T + x\mathbf{B}_{2,2}^T+\mathbf{B}_{3,2}^T}\\
		{{x^3}\mathbf{B}_{0,3}^T + {x^2}\mathbf{B}_{1,3}^T + x\mathbf{B}_{2,3}^T+\mathbf{B}_{3,3}^T}
		\end{array}} \right]^T}.
\end{aligned}
\end{equation}
 Then $\mathbf{C}=\mathbf{A}\mathbf{B}$ can be recovered from $\hat{\mathbf{C}}(x)=\hat{\mathbf{A}}(x)\hat{\mathbf{B}}(x)^T$ if we have the value of $\hat{\mathbf{C}}(x)$ for seven distinct values of $x\in\mathbb{R}$. More precisely, each entry of $\hat{\mathbf{C}}(x)$ is a polynomial of degree six. Let us focus on block $(k, \ell)$ of $\hat{\mathbf{C}}(x)$ denoted by $\hat{\mathbf{C}}_{k,\ell}(x)$. Then
\begin{equation}
\hat{\mathbf{C}}_{k,\ell}(x)=\hat{\mathbf{C}}_0^{(k,\ell)}+\hat{\mathbf{C}}_1^{(k,\ell)}x +... + \hat{\mathbf{C}}_6^{(k,\ell)}x^6.
\end{equation}
In this expansion, one can verify that $\hat{\mathbf{C}}_3^{(k,\ell)}=\mathbf{C}_{k,\ell}$. 
If we have the value of $\hat{\mathbf{C}}_{k,\ell}(x)$ for seven distinct values of $x\in \mathbb{R}$ then all the coefficients of $\hat{\mathbf{C}}_{k,\ell}(x)$ can be calculated using polynomial interpolation. In particular, $\mathbf{C}_{k,\ell}$, which is the coefficient of $x^3$, can be calculated. \\

\item
\textit{\textbf{Step 2}}. To reduce the dimension of this product,  the count-sketch method is used. Assume we construct three count-sketches for $\hat{\mathbf{A}}(x)$.
 Let us assume that the sketch polynomials of the  row-blocks of $\hat{\mathbf{A}}(x)$  are described as:
\begin{align}
&{\mathbf{F}_{1}}(x,\alpha) = \big(-{\hat{\mathbf{A}}_{0,0}}(x) + {\hat{\mathbf{A}}_{1,0}}(x)+{\hat{\mathbf{A}}_{3,0}}(x)\big) - {\hat{\mathbf{A}}_{2,0}}(x)\alpha,\\
&\mathbf{F}_{2}(x,\alpha) = \big({\hat{\mathbf{A}}_{1,0}}(x) +{\hat{\mathbf{A}}_{2,0}}(x)\big)+ \big({\hat{\mathbf{A}}_{0,0}}(x) - {\hat{\mathbf{A}}_{3,0}}(x)\big)\alpha,\\
&\mathbf{F}_{3}(x,\alpha) = {\hat{\mathbf{A}}_{2,0}}(x)+ \big(-{\hat{\mathbf{A}}_{0,0}}(x) +{\hat{\mathbf{A}}_{1,0}}(x)+ {\hat{\mathbf{A}}_{3,0}}(x)\big)\alpha,
\label{eq3}
\end{align}
 where $\mathbf{F}_i$ is the sketch polynomial of matrix $\hat{\mathbf{A}}(x)$ using the hash function $h_{i,\hat{\mathbf{A}}}:[0:3]\to[0:1]$. In particular, the hash functions are considered as follows
 \begin{align}
 	h_{1,\hat{\mathbf{A}}}(y)&=\big((y+2)\mod 3\big)\mod 2,\\
 	h_{2,\hat{\mathbf{A}}}(y)&=\big((y+1)\mod 3\big)\mod 2,\\
 	h_{3,\hat{\mathbf{A}}}(y)&=\big((8y+1)\mod 11\big)\mod 2.
 \end{align}

 Same as before, assume we have three count-sketches for $\hat{\mathbf{B}}(x)$. To be specific, assume that the related sketch polynomials are defined as follows 
\begin{align}\label{eq30}
&{\mathbf{G}_{1}}(x,\alpha) = \big({-\hat{\mathbf{B}}_{0,1}}(x) -\hat{\mathbf{B}}_{0,3}(x)\big) +\big({\hat{\mathbf{B}}_{0,0}}(x) + {\hat{\mathbf{B}}_{0,2}}(x)\big)\alpha, \\
&\mathbf{G}_{2}(x,\alpha) = \big({{\hat{\mathbf{B}}}_{0,0}}(x) -{\hat{\mathbf{B}}_{0,1}}(x)\big)+ \big(-{{\hat{\mathbf{B}}}_{0,2}}(x) + {{\hat{\mathbf{B}}}_{0,3}}(x)\big)\alpha,\\
&\mathbf{G}_{3}(x,\alpha) = \big(-{{\hat{\mathbf{B}}}_{0,1}}(x) -{\hat{\mathbf{B}}_{0,2}}(x)+{\hat{\mathbf{B}}_{0,3}}(x)\big)+ {{\hat{\mathbf{B}}}_{0,0}}(x)\alpha,
\end{align}
 where $\mathbf{G}_i$ is the sketch polynomial of matrix $\hat{\mathbf{B}}(x)$ using the hash function $h_{i,\hat{\mathbf{B}}}:[0:3]\to[0:1]$ as follows
\begin{align}
h_{1,\hat{\mathbf{B}}}(y)&=(y+1)\mod 2,\\
h_{2,\hat{\mathbf{B}}}(y)&=\big((3y+2)\mod 5\big)\mod 2,\\
h_{3,\hat{\mathbf{B}}}(y)&=\big((2y+3)\mod 5\big)\mod 2.
\end{align}
We note that the ${\mathbf{F}_{\ell}}(x,\alpha){\mathbf{G}_{\ell}}(x,\alpha)^T$ can be considered as a sketch polynomial for $\hat{\mathbf{C}}(x)$, where $\ell=1,2,3$. For example, ${\mathbf{F}_{1}}(x,\alpha){\mathbf{G}_{1}}(x,\alpha)^T$  can be written as
\begin{equation}\label{eqP}
\mathbf{P}_{1,1}(x,\alpha)\triangleq{\mathbf{F}_{1}}(x,\alpha){\mathbf{G}_{1}}(x,\alpha)^T=\sum_{i=0}^{2}{{\mathbf{P}}^{(1,1)}_{i}(x)\alpha^i},
\end{equation}
where in this expansion,
\begin{align}
&{\mathbf{P}}^{(1,1)}_{0}(x)\hspace{-1mm}=\hspace{-1mm}\hat{\mathbf{C}}_{0,1}(x)+\hat{\mathbf{C}}_{0,3}(x)-\hat{\mathbf{C}}_{1,1}(x)-\hat{\mathbf{C}}_{1,3}(x)-\hat{\mathbf{C}}_{3,1}(x)-\hat{\mathbf{C}}_{3,3}(x),\\
&{\mathbf{P}}^{(1,1)}_{1}(x)\hspace{-1mm}=\hspace{-1mm}-\hat{\mathbf{C}}_{0,0}(x)-\hat{\mathbf{C}}_{0,2}(x)+\hat{\mathbf{C}}_{1,0}(x)+\hat{\mathbf{C}}_{1,2}(x)+\hat{\mathbf{C}}_{2,1}(x)+\hat{\mathbf{C}}_{2,3}(x)+\hat{\mathbf{C}}_{3,2}(x)+\hat{\mathbf{C}}_{3,0}(x),\\
&{\mathbf{P}}^{(1,1)}_{2}(x)\hspace{-1mm}=\hspace{-1mm}-\hat{\mathbf{C}}_{2,0}(x)-\hat{\mathbf{C}}_{2,2}(x).
\end{align} 
  Each entry of $\hat{\mathbf{C}}(x)$ is a polynomial of degree six in which the coefficient of $x^3$ is the combination of entries of original result $\mathbf{C}$. This can be explained better as follows
\begin{equation}\label{eq34}
\mathbf{P}_{1,1}(x,\alpha)=\sum_{i=0}^{2}\sum_{j=0}^{6}{\mathbf{P}^{(1,1)}_{ij}x^j \alpha^i}.
\end{equation}
Then according to \eqref{23} and \eqref{24}, and the discussion followed we have 
\begin{align}\label{Ps}
&\mathbf{P}^{(1,1)}_{03}=\mathbf{C}_{0,1}+\mathbf{C}_{0,3}-\mathbf{C}_{1,1}-\mathbf{C}_{1,3}-\mathbf{C}_{3,1}-\mathbf{C}_{3,3},\\\label{Ps2}
&\mathbf{P}^{(1,1)}_{13}=-\mathbf{C}_{0,0}-\mathbf{C}_{0,2}+\mathbf{C}_{1,0}+\mathbf{C}_{1,2}+\mathbf{C}_{2,1}+\mathbf{C}_{2,3}+\mathbf{C}_{3,2}+\mathbf{C}_{3,0},\\
\label{Ps3}
&\mathbf{P}^{(1,1)}_{23}=-\mathbf{C}_{2,0}-\mathbf{C}_{2,2}.
\end{align}  
Particularly in the expansion \eqref{eq34}, the terms in \eqref{Ps}--\eqref{Ps3} are of interest. The reason the other coefficients are not interesting is that the coefficients that we are looking for appear only in these terms. Thus, we have another count-sketch hidden in the count-sketch of $\hat{\mathbf{C}}(x)$. These three coefficients \eqref{Ps}--\eqref{Ps3} form a count-sketch of the blocks of $\mathbf{C}$. The reason that the special coefficients of \eqref{eq34} form a count-sketch of the blocks of $\mathbf{C}$, is the structure used in this scheme. This structure, which we name \emph{CodedSketch}, is the concatenation of the count-sketch and the entangled polynomial code. Other ${\mathbf{F}_{\ell}}(x,\alpha){\mathbf{G}_{\ell}}(x,\alpha)^T$ form independent count-sketch similarly.\\
 In the following, the computation of these sketch polynomials over a distributed system is proposed and
 to form the results of ${\mathbf{F}_{\ell}}(x,\alpha){\mathbf{G}_{\ell}}(x,\alpha)^T$ where $\ell=1,2,3$ efficiently, we use another layer of structured code motivated by improved entangled polynomial codes~\cite{entangle}.\\

\item
 \textit{\textbf{Step 3}}.	The master node creates the following polynomials and encodes the $\mathbf{F}_\ell(x,\alpha)$ and $\mathbf{G}_\ell(x,\alpha)$ 
 	\begin{align}
 	 &	\mathbf{F}(x,\alpha,\omega) \triangleq\mathbf{F}_{1}(x,\alpha)(\frac{\omega-\beta_2}{\beta_1-\beta_2})(\frac{\omega-\beta_3}{\beta_1-\beta_3}) + \mathbf{F}_{2}(x,\alpha)(\frac{\omega-\beta_1}{\beta_2-\beta_1})(\frac{\omega-\beta_3}{\beta_2-\beta_3})+\mathbf{F}_{3}(x,\alpha)(\frac{\omega-\beta_1}{\beta_3-\beta_1})(\frac{\omega-\beta_2}{\beta_3-\beta_2}),\\
 	 	&	\mathbf{G}(x,\alpha,\omega) \triangleq\mathbf{G}_{1}(x,\alpha)(\frac{\omega-\beta_2}{\beta_1-\beta_2})(\frac{\omega-\beta_3}{\beta_1-\beta_3}) + \mathbf{G}_{2}(x,\alpha)(\frac{\omega-\beta_1}{\beta_2-\beta_1})(\frac{\omega-\beta_3}{\beta_2-\beta_3})+\mathbf{G}_{3}(x,\alpha)(\frac{\omega-\beta_1}{\beta_3-\beta_1})(\frac{\omega-\beta_2}{\beta_3-\beta_2}),
 	\label{eq6}
 	\end{align}
 	where $\beta_1,\beta_2,\beta_3\in\mathbb{R}$ are three distinct values.
  These polynomials are linear combinations of  sketch polynomials created using different hash functions. It can be seen that $\mathbf{F}(x,\alpha,\beta_1)=\mathbf{F}_1(x,\alpha)$,  $\mathbf{F}(x,\alpha,\beta_2)=\mathbf{F}_2(x,\alpha)$ and $\mathbf{F}(x,\alpha,\beta_3)=\mathbf{F}_3(x,\alpha)$ and so does $\mathbf{G}(x,\alpha,\omega)$.  
   Since the extraction of hidden count-sketch of the blocks of $\mathbf{C}$ is desired, we choose $\alpha=x^4$ and $\omega=x^{15}$.   Let 
  $\mathbf{F}(x)\triangleq\mathbf{F}(x,x^4,x^{15})$ and $\mathbf{G}(x)\triangleq\mathbf{G}(x,x^4,x^{15})$. The number $\theta_j\in \mathbb{R}$ is dedicated to the $j$th worker node, where $\theta_i\ne \theta_j$ if $i \ne j$.
Therefore, the master node sends $\mathbf{F}(\theta_j)\in \mathbb{R}^{\frac{r}{4}\times \frac{s}{4}}$ and $\mathbf{G}(\theta_j)\in \mathbb{R}^{\frac{s}{4}\times \frac{t}{4}}$ to $j$th worker node.\\

\item
\textit{\textbf{Step 4}.}
 	Having received matrices $\mathbf{F}(\theta_j)$ and $\mathbf{G}(\theta_j)$ from master node, the $j$th worker node multiplies these two matrices.
 	  Then it returns the result, i.e., $\mathbf{F}(\theta_j)\mathbf{G}(\theta_j)^T$ to the master node. The result calculated at node $j$ can be written as
 	\begin{align}
 	\begin{aligned}
	\mathbf{F}(\theta_j)\mathbf{G}(\theta_j)^T&=
	\big(\mathbf{F}_1(\theta_j)\mathbf{G}_1(\theta_j)^T\big)(\frac{\omega-\beta_2}{\beta_1-\beta_2})^2(\frac{\omega-\beta_3}{\beta_1-\beta_3})^2\\ &+\big(\mathbf{F}_2(\theta_j) \mathbf{G}_2(\theta_j)^T\big)(\frac{\omega-\beta_1}{\beta_2-\beta_1})^2(\frac{\omega-\beta_3}{\beta_2-\beta_3})^2\\
	&+\big(\mathbf{F}_3(\theta_j) \mathbf{G}_3(\theta_j)^T\big)(\frac{\omega-\beta_1}{\beta_3-\beta_1})^2(\frac{\omega-\beta_2}{\beta_3-\beta_2})^2\\
	&+\sum\limits_{\ell = 1}^3 {\sum\limits_{k= 1\hfill\atop k\ne \ell\hfill}^3 {\bigg[{\mathbf{F}_\ell}({\theta _j}){\mathbf{G}_k}(} } {\theta _j}{)^T}(\prod\limits_{\scriptstyle{i_1} = 1\hfill\atop
		\scriptstyle{i_1} \ne \ell\hfill}^3 {(\frac{{\omega - \beta_{i_1}}}{{\beta_\ell - \beta_{i_1}}})} )(\prod\limits_{\scriptstyle{i_2} = 1\hfill\atop
		\scriptstyle{i_2} \ne k\hfill}^3 {(\frac{{\omega - \beta_{i_2}}}{{\beta_k - \beta_{i_2}}})} )\bigg]
	\label{eq15}
	\end{aligned}
 	\end{align}
%
 	 By substituting the $\mathbf{F}_\ell$ and $\mathbf{G}_k$ in \eqref{eq15}, where $\ell,k=1,2,3$, the polynomial with $75$ coefficients is created in which the count-sketch results of the blocks of $\mathbf{C}$ are located.\\
 	 
 \item	 
 \textit{\textbf{Step 5}.}
 The master node can recover all of the polynomials' coefficients by receiving the computation results of any 75 worker nodes. That is because the recovering process is equivalent to interpolating a polynomial of degree 74 given its value at 75 points. After interpolation and recovering the coefficients, a polynomial of degree 74 is created. Assume that, in this polynomial, all $x^{15}$ are replaced by $\omega$. In this case, a bivariate polynomial is achieved in which can we choose $\omega=\beta_1,\beta_2,\beta_3$ to calculate three sketch polynomials $\mathbf{F}_1(x)\mathbf{G}_1(x)^T$, $\mathbf{F}_2(x) \mathbf{G}_2(x)^T$ and $\mathbf{F}_3(x) \mathbf{G}_3(x)^T$ for $\hat{\mathbf{C}}(x)$ respectively. According to \eqref{Ps} we need the coefficients of $x^3$, $x^7$ and $x^{11}$  to find the hidden count-sketches of the blocks of $\mathbf{C}$ in the sketch polynomial of $\hat{\mathbf{C}}(x)$.\\
\item
\textit{\textbf{Step 6.}}
 The coefficients $x^3$, $x^7$ and $x^{11}$ of the three sketch polynomials $\mathbf{F}_1(x)\mathbf{G}_1(x)^T$, $\mathbf{F}_2(x) \mathbf{G}_2(x)^T$ and $\mathbf{F}_3(x) \mathbf{G}_3(x)^T$  are shown in \tabref{table}.  To achieve an approximation of ${\mathbf{C}}$ more accurately, the master node takes the median of these estimations after multiplying them to the corresponding sign functions. For example, to approximate the value of $[\mathbf{C}_{2,0}]_{i,j}$ for all $i\in[0:\frac{r}{m}-1]$ and $j\in[0:\frac{t}{n}-1]$, the master node does the following
 \begin{equation}
[\mathbf{C}_{2,0}]_{i,j} \approx \text{median}\bigg\{s_1(2)\hat{s}_1(0)[\mathbf{P}_{23}^{(1,1)}]_{i,j},\hspace{0.3cm}
s_2(2)\hat{s}_2(0)[\mathbf{P}_{03}^{(2,2)}]_{i,j},\hspace{0.3cm}
s_3(2)\hat{s}_3(0)[\mathbf{P}_{13}^{(3,3)}]_{i,j}
\bigg\}.
\end{equation}
 In another form the master node takes the median of the following terms
 \begin{equation}
  \begin{aligned}
  &[\mathbf{C}_{2,0}]_{i,j} \approx \text{median}\bigg\{[\mathbf{C}_{2,0}]_{i,j}+[\mathbf{C}_{2,2}]_{i,j},\hspace{0.3cm}
 	[\mathbf{C}_{1,0}]_{i,j}-[\mathbf{C}_{1,1}]_{i,j}+[\mathbf{C}_{2,0}]_{i,j}-[\mathbf{C}_{2,1}]_{i,j},\\
  &[\mathbf{C}_{0,1}]_{i,j}+[\mathbf{C}_{0,2}]_{i,j}-[\mathbf{C}_{0,3}]_{i,j}-[\mathbf{C}_{1,1}]_{i,j}-[\mathbf{C}_{1,2}]_{i,j}+[\mathbf{C}_{1,3}]_{i,j}+[\mathbf{C}_{2,0}]_{i,j}-[\mathbf{C}_{3,1}]_{i,j}-[\mathbf{C}_{3,2}]_{i,j}+[\mathbf{C}_{3,3}]_{i,j}
  \bigg\}.
  \end{aligned}
 \end{equation}
\begin{table}[htp]
	\centering \caption{\footnotesize The Count-Sketch results of $\mathbf{C}$}
	\renewcommand*{\arraystretch}{2}
	\begin{tabular}{| c || c | c| c| }
		\hline 
		Index & 0 & 1& 2 \\
		\hline\hline
		1st Count-Sketch & $\begin{aligned}[t]
		&\mathbf{P}_{03}^{(1,1)}=\mathbf{C}_{0,1}+\mathbf{C}_{0,3}-\mathbf{C}_{1,1}\\&-\mathbf{C}_{1,3}-\mathbf{C}_{3,1}-\mathbf{C}_{3,3}\end{aligned}$ & $\begin{aligned}[t]
		&\mathbf{P}_{13}^{(1,1)}=-\mathbf{C}_{0,0}-\mathbf{C}_{0,2}+\mathbf{C}_{1,0}\\&+\mathbf{C}_{1,2}+\mathbf{C}_{2,1}+\mathbf{C}_{2,3}+\mathbf{C}_{3,2}+\mathbf{C}_{3,0}\end{aligned}$ & $\mathbf{P}_{23}^{(1,1)}=-\mathbf{C}_{2,0}-\mathbf{C}_{2,2}$
		\\
		\hline
		2nd Count-Sketch &  $\begin{aligned}[t]
		&\mathbf{P}_{03}^{(2,2)}=\mathbf{C}_{1,0}-\mathbf{C}_{1,1}+\mathbf{C}_{2,0}\\&-\mathbf{C}_{2,1}\end{aligned}$ & $\begin{aligned}[t]
		&\mathbf{P}_{13}^{(2,2)}=\mathbf{C}_{0,0}-\mathbf{C}_{0,1}-\mathbf{C}_{1,2}+\mathbf{C}_{1,3}\\&-\mathbf{C}_{2,2}+\mathbf{C}_{2,3}-\mathbf{C}_{3,0}+\mathbf{C}_{3,1}\end{aligned}$ & $\begin{aligned}[t]
		&\mathbf{P}_{23}^{(2,2)}=-\mathbf{C}_{0,2}+\mathbf{C}_{0,3}\\&+\mathbf{C}_{3,2}-\mathbf{C}_{3,3}\end{aligned}$ 
		\\
		\hline
		3rd Count-Sketch &  $\begin{aligned}[t]
		\mathbf{P}_{03}^{(3,3)}=-\mathbf{C}_{2,1}-\mathbf{C}_{2,2}+\mathbf{C}_{2,3}\end{aligned}$ & $\begin{aligned}[t]
		&\mathbf{P}_{13}^{(3,3)}=\mathbf{C}_{0,1}+\mathbf{C}_{0,2}-\mathbf{C}_{0,3}-\mathbf{C}_{1,1}\\&-\mathbf{C}_{1,2}+\mathbf{C}_{1,3}+\mathbf{C}_{2,0}-\mathbf{C}_{3,1}-\mathbf{C}_{3,2}\\&+\mathbf{C}_{3,3}\end{aligned}$ & $\begin{aligned}[t]
		&\mathbf{P}_{23}^{(3,3)}=-\mathbf{C}_{0,0}+\mathbf{C}_{1,0}\\&+\mathbf{C}_{3,0}\end{aligned}$ 
		\\
		\hline
	\end{tabular}
	\label{table}
\end{table}

\end{itemize}
  Therefore, in this scheme, the master node sends two matrices  to each worker and each worker sends back the result of multiplication of these matrices to the master node. If a sufficient subset of worker nodes return their results to the master, finally the master node can recover an approximation of $\mathbf{C}$ by using polynomial interpolation and median of the desired coefficients.
%
\subsection{General CodedSketch Design}
Now we present the proposed scheme for distributed computation of an approximation for the matrix multiplication using CodedSketch in a general setting. In summary, the encoding step of CodedSketch has three important parts which use the ideas of entangled polynomial codes and count-sketch to form the useful combination of randomness and structured codes such that an approximated version of multiplication is computed. First of all, the input matrices $\mathbf{A}$ and $\mathbf{B}$ are partitioned into equal-size $m\times p$ and equal-size $p\times n$ sub-matrices as \eqref{eq7} and \eqref{eq8}.  In this setting, the following steps will be taken:

\begin{itemize}
	\item \textit{\textbf{Step 1:}}
The master node forms the following polynomial coded matrices based on column-blocks of $\mathbf{A}$ and row-blocks of $\mathbf{B}$ as follows
	\begin{align}\label{49}
	\hat{\mathbf{A}}(x)&=\sum_{k=0}^{p-1}{x^k\mathbf{A}_k},\\
	\label{50}
	\hat{\mathbf{B}}(x)^T&=\sum_{k=0}^{p-1}{x^{p-1-k}\mathbf{B}_k^T},
	\end{align}
	 where $\mathbf{A}_k$ and $\mathbf{B}_k^T$ are the $k$th column of $\mathbf{A}$ and $k$th row of $\mathbf{B}$, for $k=0,\ldots, p-1$. By this coding, one can verify that the block $(k,\ell)$ of $\hat{\mathbf{C}}(x)=\hat{\mathbf{A}}(x)\hat{\mathbf{B}}(x)^T$ is a polynomial of degree $(2p-2)$, where the coefficient of $x^{p-1}$ is the block $(k, \ell)$ of $\mathbf{C}$. More precisely, the block $(k, \ell)$ of $\hat{\mathbf{C}}(x)$ can be written as 
	\begin{equation}
	 \hat{\mathbf{C}}_{k,\ell}(x)=\hat{\mathbf{C}}_0^{(k,\ell)}+\hat{\mathbf{C}}_1^{(k,\ell)}x +... + \hat{\mathbf{C}}_{2p-2}^{(k,\ell)}x^{2p-2},
	 \end{equation}
	 where  in particular $\hat{\mathbf{C}}_{p-1}^{(k,\ell)}=\mathbf{C}_{k,\ell}$, and  $\hat{\mathbf{C}}_{\nu}^{(k,\ell)}$, $\nu=0,\ldots, 2p-2$, $\nu \neq p-1$ are some real  matrices with the same size as $\hat{\mathbf{C}}_{p-1}^{(k,\ell)}$.
	This means that if we have the value of $\hat{\mathbf{C}}(x)$ for $2p-1$ distinct values of $x\in \mathbb{R}$, we can recover $\hat{\mathbf{C}}(x)$, and in particular the result $\mathbf{C}$.  However, 
our goal is not to compute $\mathbf{C}$ exactly. In the following, we use count-sketch method, to compute the approximation of $\mathbf{C}$, and in return reduce the recovery threshold.

	\item \textit{\textbf{Step 2:}}
In this step, we use the hash functions $h_i:\mathcal{U}\to[0:b'-1]$ and $\tilde{h}_i:\mathcal{U}\to[0:b'-1]$,   chosen uniformly at random from the family of 2-wise independent hash functions to sketch the row-blocks and the column-blocks of coded matrices $\hat{\mathbf{A}}(x)$ and $\hat{\mathbf{B}}(x)^T$ respectively, for   $i=1,...,d$, and for some $b' \in \mathbb{N}$.  
To sketch, we also need $s_{\ell}:\mathcal{U} \to \{-1,1\}$ and $\tilde{s}_{\ell}:\mathcal{U}\to\{-1,1\}$, for   $\ell=1,\ldots,d$,  which are chosen uniformly at random  from the family of pairwise--independent sign functions.  Let $\mathbf{F}_{\ell}(x,\alpha)$ and $\mathbf{G}_{\ell}(x, \alpha)^T$ be the sketch polynomials of the row-blocks of $\hat{\mathbf{A}}(x)$ based on $h_{\ell}$ and $s_{\ell}$,   and the column-blocks of $\hat{\mathbf{B}}(x)^T$ based on $\tilde{h}_{\ell}$ and $\tilde{s}_{\ell}$, for $\ell=1,\ldots,d$.  More precisely, 
\begin{align}\label{52}
	&\mathbf{F}_{\ell}(x,\alpha)=\sum_{i=0}^{m-1}{s_{\ell}(i)\hat{\mathbf{A}}_{i,0}(x)\alpha^{h_{\ell}(i)}},\\
	\label{53}
&\mathbf{G}_{\ell}(x,\alpha)=\sum_{j=0}^{n-1}{\tilde{s}_{\ell}(j)\hat{\mathbf{B}}_{0,j}(x)\alpha^{\tilde{h}_{\ell}(j)}},
	\end{align}
	where $\hat{\mathbf{A}}_{i,0}(x)$ and $\hat{\mathbf{B}}_{0,i}(x)$ are the $i$th block element of $\hat{\mathbf{A}}(x)$ and $\hat{\mathbf{B}}(x)$. 
	
	\item  \textit{\textbf{Step 3:}}
	The master node creates the following polynomials using improved entangled polynomial codes
	\begin{align}
	\label{eq54}
&	\mathbf{F}(x,\alpha,\omega) = \sum\limits_{\ell=1}^{d}\big({\mathbf{F}_{\ell}(x,\alpha)\prod\limits_{\scriptstyle{i} = 1\hfill\atop
		\scriptstyle{i} \ne \ell\hfill}^d {(\frac{{\omega - \beta_{i}}}{{\beta_\ell -\beta_{i}}})}}\big), \\
			\label{eq55}		
&	\mathbf{G}(x,\alpha, \omega) = \sum\limits_{\ell=1}^{d}\big({\mathbf{G}_{\ell}(x,\alpha)\prod\limits_{\scriptstyle{i} = 1\hfill\atop
			\scriptstyle{i} \ne \ell\hfill}^d {(\frac{{\omega - \beta_{i}}}{{\beta_\ell - \beta_{i}}})}}\big),
	\end{align}
	for some variables $\alpha$ and $\omega$, where $\beta_1,\dots,\beta_d\in\mathbb{R}$ are some distinct values. 
	
	The master node chooses $\alpha=x^p$ and $\omega=x^{2pb'-1}$. Let $\mathbf{F}(x) \buildrel \Delta \over =\mathbf{F}(x,x^p,x^{2pb'-1})$ and  $\mathbf{G}(x) \buildrel \Delta \over =\mathbf{G}(x,x^p,x^{2pb'-1})$. Then, the master node sends $\mathbf{F}(\theta_j)$ and $\mathbf{G}(\theta_j)$ matrices to the $j$th worker node, where $\theta_j$ is an arbitrary element in $\mathbb{R}$ and  $\theta_i\ne \theta_j$ if $i\ne j$. The size of the transmitted matrices are $\frac{1}{pm}$ fraction of $\mathbf{A}$ and $\frac{1}{pn}$ fraction of $\mathbf{B}$ respectively. In other words, $\mathbf{F}(\theta_j)\in \mathbb{R}^{\frac{r}{m}\times\frac{s}{p}}$ and $\mathbf{G}(\theta_j)\in \mathbb{R}^{\frac{s}{p}\times\frac{t}{n}}$.


	
	\item \textit{\textbf{Step 4:}}
	Worker node $j$ has $\mathbf{F}(\theta_j)$ and $\mathbf{G}(\theta_j)$  and computes $\mathbf{P}({\theta_j})\buildrel \Delta \over =\mathbf{F}(\theta_j)\mathbf{G}(\theta_j)^T$, and sends the result to the master node.

\item \textit{\textbf{Step 5:}}
	The master node waits to receive answers from $(2pb'-1)(2d-1)$ worker nodes. It can be verified that the polynomial
	\begin{equation}
	\label{eq56}
\mathbf{P}(x)\buildrel \Delta \over =\mathbf{F}(x)\mathbf{G}(x)^T= \sum\limits_{\ell = 1}^d {\sum\limits_{k= 1}^d {\bigg[{\mathbf{F}_\ell}({x}){\mathbf{G}_k}(}} {x}{)^T}(\prod\limits_{\scriptstyle{i_1} = 1\hfill\atop
	\scriptstyle{i_1} \ne \ell\hfill}^d {(\frac{{\omega - \beta_{i_1}}}{{\beta_\ell -\beta_{i_1}}})} )(\prod\limits_{\scriptstyle{i_2} = 1\hfill\atop
	\scriptstyle{i_2} \ne k\hfill}^d {(\frac{{\omega - \beta_{i_2}}}{{\beta_k -\beta_{i_2}}})} )\bigg],
	\end{equation}
	 has degree of $(2pb'-1)(2d-1)-1$, where $\omega=x^{2pb'-1}$. Thus, the master node  can interpolate $\mathbf{P}(x)$, having access to $\mathbf{P}(\theta_j)$, for $(2pb'-1)(2d-1)$ distinct values of $\theta_j$.

Assume that, in the polynomial $\mathbf{P}(x)$, the master node replaces $x^{2pb'-1}$ by variable $\omega$. 
By this substitution a bivariate polynomial of $x$ and $\omega$ is formed. To recover $\mathbf{F}_{\eta}(x) \mathbf{G}_{\eta}(x)^T$, for any $\eta \in \{1, \ldots, d \}$, the master node then replaces the variable $\omega$ with $\beta_{\eta}$.

In Appendix \ref{appC}, we show that in $\mathbf{F}_{\eta}(x) \mathbf{G}_{\eta}(x)^T$, the coefficients of  $x^{kp+p-1}$ for $k=0,...,2b'-2$, denoted by $\mathbf{P}_{k,p-1}^{(\eta,\eta)}$, are equal to
\begin{equation}\label{57}
\mathbf{P}_{k,p-1}^{(\eta,\eta)}=\hspace{-0.5cm}\sum_{i,j: h_\eta(i)+\tilde{h}_\eta(j)=k}\hspace{-0.5cm}{s_\eta(i)\tilde{s}_\eta(j)\mathbf{C}_{i,j}},
\end{equation}
	where $\mathbf{C}_{i,j}$ is the block $(i,j)$ of $\mathbf{C}$.
	\item \textit{\textbf{Step 6:}}
	Recall that the goal is to approximate the original result of multiplication, i.e., $\mathbf{C}$.
	\begin{equation}
	\tilde{\mathbf{C}} = \left[ {\begin{array}{*{20}{c}}
		{{\tilde{\mathbf{C}} _{0,0}}}&{{\tilde{\mathbf{C}} _{0,1}}}&{...}&{{\tilde{\mathbf{C}} _{0,n - 1}}}\\
		{{\tilde{\mathbf{C}} _{1,0}}}&{{\tilde{\mathbf{C}} _{1,1}}}&{...}&{{\tilde{\mathbf{C}} _{1,n - 1}}}\\
		\vdots & \vdots & \ddots & \vdots \\
		{{\tilde{\mathbf{C}} _{m - 1,0}}}&{{\tilde{\mathbf{C}} _{m - 1,1}}}&{...}&{{\tilde{\mathbf{C}} _{m - 1,n - 1}}}
		\end{array}} \right]\approx\mathbf{C}.
	\end{equation}
	
%
	For each $[\tilde{\mathbf{C}}_{k,\ell}]_{i,j}$, the recovering phase is done as follows
	\begin{equation}
	[\tilde{\mathbf{C}}_{k,\ell}]_{i,j}=\text{median}\{s_{\eta}(k)\tilde{s}_{\eta}(\ell)\hspace{0.1cm}\big[\mathbf{P}_{{h_{\eta}(k)+\tilde{h}_{\eta}(\ell)},p-1}^{(\eta,\eta)}\big]_{i,j} \}^{d}_{\eta=1},
	\end{equation}
where $i\in[0:\frac{r}{m}-1]$, $j\in[0:\frac{t}{n}-1]$, $k\in[0,m-1]$ and $\ell\in[0,n-1]$.

	\end{itemize}

	
%

 \begin{proof}[\textbf{Proof of Theorem \ref{th3}}]\label{proof4}
 We note that  each  count-sketch used for $\bA$ and $\bB$ in~\eqref{52} and \eqref{53} has length $b'$, while each resulting count-sketch in~\eqref{57} for $\mathbf{C}$ has length $b=2b'-1$. 
 
 According to Theorem \ref{Th1} in Section~\ref{ProbFor} on preliminaries, to have an unbiased, $(\epsilon, \delta)$-accurate approximation of $\textbf{C}$,  it is sufficient to choose $d\geq \log \frac{1}{\delta }$ and  $b' \geq \frac{3}{\epsilon ^2}$. On the other hand, from~\eqref{eq56}, the degree of the polynomial $\mathbf{P}(x)$ is equal to $\big((2pb'-1)(2d-1)-1\big)$. Thus, to recover  $\mathbf{P}(x)$, the master node needs $\mathbf{P}(\theta_j)$ for $\big((2pb'-1)(2d-1)\big)$ distinct $\theta_j$'s. 
 Thus the recovery threshold of the scheme is the minimum of $\big((2pb'-1)(2d-1)\big)$ subject to  $d\geq \log \frac{1}{\delta }$, $b' \geq \frac{3}{\epsilon ^2}$, and  $d, b'\in \mathbb{N}$, as follows
		\begin{equation}
		N^{*}(\epsilon,\delta,p,m,n) \le \min\left\{ {(2p\left\lceil {\frac{3}{{{\epsilon ^2}}}} \right\rceil  - 1)(2\left\lceil {\log \frac{1}{\delta }} \right\rceil  - 1) - 1}, pmn+p-1 \right\}.
		\end{equation}
	
 \end{proof}
  \begin{proof}[\textbf{Proof of Theorem \ref{th4}}]\label{proof5}
		According to Theorem \ref{th2} in Section~\ref{ProbFor} on preliminaries, if we choose the length of the count-sketches as $b'=\mathcal{O}(\frac{\kappa}{\epsilon^2})$ and the number of count-sketches as $d$, then we have the following guarantee 
		\begin{align}\label{errK}
		\mathbf{P}\bigg[|[\mathbf{C}_{k,\ell}]_{i,j}-[\tilde{\mathbf{C}}_{k,\ell}]_{i,j}|\ge\frac{\epsilon}{\sqrt{\kappa}}\mathrm{err}_2^{(\kappa,(i,j))}(\mathbf{C}) \bigg]\le2e^{-\Omega(d\epsilon^2)}\hspace{1cm} \forall i,j,k,\ell,
		\end{align}
		where $\mathrm{err}_2^{(\kappa,(i,j))}(\mathbf{C})={(\sum_{(k,\ell)\notin \mathcal{T}_{\kappa}}{|[\mathbf{C}_{k,\ell}]_{i,j}|^2})}^\frac{1}{2}$
		where $\mathcal{T}_{\kappa}$ is the set of
		indices of the blocks of $\mathbf{C}$ whose entry $(i,j)$ is in the $\kappa$-largest entry $(i,j)$ among all blocks. Also, $i\in[0:\frac{r}{m}-1]$, $j\in[0:\frac{t}{n}-1]$, $k\in[0:m-1]$ and $\ell\in[0:n-1]$. From \eqref{errK} we have the following guarantee
		\begin{align}
		\mathbf{P}\bigg[|[\mathbf{C}]_{i,j}-[\tilde{\mathbf{C}}]_{i,j}|\ge\frac{\epsilon}{\sqrt{\kappa}}\mathrm{err}_2^{(\kappa,(i',j'))}(\mathbf{C}) \bigg]\le2e^{-\Omega(d\epsilon^2)},\hspace{1cm}\forall i\in[0:r-1], j\in[0:t-1],
		\end{align} 
		where $i'=i \mod \frac{r}{m}$, $j'=j \mod \frac{t}{n}$, and $[\mathbf{C}]_{i,j}$ is the $(i,j)$th entry of $\mathbf{C}$. \\Recall that 
		$\mathrm{err}_2^{(\kappa,(i',j'))}(\mathbf{C})={(\sum_{(k,\ell)\notin \tilde{\mathcal{T}}_{\kappa}}{|[\mathbf{C}_{k,\ell}]_{i',j'}|^2})}^\frac{1}{2}$,
		where $\tilde{\mathcal{T}}_{\kappa}$ is the set of
		indices of the blocks of $\mathbf{C}$ whose entry $(i',j')$ is in the $\kappa$-largest entry $(i',j')$ among all blocks.
		Thus, according to Remark \ref{remark3}, and the result of Theorem \ref{th3}, it is sufficient to choose $b'\ge \frac{3\kappa}{\epsilon^2}$ and $d\ge\log(mn)$, where $mn$ is the number of block entries of $\mathbf{C}$.	
		 So, if the result of multiplication is $\kappa$-sparse, the CodedSketch method computes $\mathbf{C}$ with probability $1\hspace{-0.5mm}-\hspace{-0.5mm}\delta\hspace{-0.5mm}=\hspace{-0.5mm}1\hspace{-0.5mm}-\hspace{-0.5mm}\frac{1}{(mn)^{\Theta(1)}}$ by choosing parameters properly. So, we have
			\begin{equation}
			N^{*}(\epsilon,\delta,p,m,n) \le \min\left\{(2p\left\lceil {\frac{3\kappa}{{{\epsilon^2}}}} \right\rceil  - 1)(2\left\lceil {\log {mn} } \right\rceil  - 1) - 1, pmn+p-1 \right\}.
			\end{equation}
		
  \end{proof}
\subsection{ Complexity Analysis}\label{COMP}
\textbf{Encoding Complexity:} 
In {CodedSketch} scheme, the master node in the encoding step performs three steps:\\
1- The master node forms the polynomial matrices using entangled polynomial codes based on the column-blocks of $\mathbf{A}$ and the row-blocks5 of $\mathbf{B}$. In other words, the master node forms two polynomial matrices in which the columns of the partitioned matrix $\mathbf{A}$ and the rows of the partitioned matrix $\mathbf{B}$ are their coefficients respectively.\\
2- The master node constructs sketch polynomials. The computational complexity of creating a sketch polynomial for a vector with $n$ entries is $\mathcal{O}(n)$. So, the computational complexity of constructing $d$ sketch polynomials for $\hat{\mathbf{A}}(x)$ and $\hat{\mathbf{B}}(x)^T$ is $\mathcal{O}(dm+dn)$. Considering the size of  each block in the partitioned matrices, the overall computational complexity in this step is $\mathcal{O}(dm\frac{r}{m}\frac{s}{p}+dn\frac{s}{p}\frac{t}{n})$.\\
3- The master node encodes $d$ sketch polynomials and creates two polynomials $\mathbf{F}(x)$ and $\mathbf{G}(x)$ using improved entangled polynomial codes.\\
4- The master node evaluates two polynomials in $N$ distinct values. Some efficient polynomial algorithms such as Horner's method \cite{pan1966means} can be used to evaluate a polynomial of degree $n$ in one point with the complexity of  $\mathcal{O}(n)$. In this phase of the algorithm, we need to compute two polynomials of degree $(2pdb'-pb'-d)$, where the coefficients are matrices of size $\frac{rs}{mp}$ and $\frac{st}{pn}$, at $N$ points. Thus, this step requires a complexity of $\mathcal{O}(Nsdb'(\frac{r}{m}+\frac{t}{n}))$.\\
Therefore, the complexity of the encoding process of CodedSketch is $\mathcal{O}(\frac{sd}{p}(r+t)+Nsdb'(\frac{r}{m}+\frac{t}{n}))$, where  $\frac{2b'}{mn}\le 1$, according to the properties of count-sketch method.

\textbf{Decoding Complexity:}
The computation performed by the master node in the decoding step includes  interpolation of a matrix-polynomial of degree $((2pb'-1)(2d-1)-1)$, where the coefficients are matrices of size $\frac{rt}{mn}$.  The complexity of interpolation of a polynomial of degree $k-1$ is $\mathcal{O}(k\log^2k)$~\cite{kung1973fast,li2000arithmetic}. 
Thus complexity of decoding in this scheme is  $\mathcal{O}(2prtd\frac{2b'}{mn}\log^2(4pb'd))$, where $\frac{2b'}{mn}\le 1$. This computation complexity is a linear function of the input size. Using this interpolation, the count-sketches of $\mathbf{C}$ are obtained. The median of these count-sketches for $rt$ times is returned as an approximation of $\mathbf{C}$ in the last step, which has a complexity of $O(rtd)$ on average
\cite{Quickselect}. 

\textbf{Computational Complexity at Each Worker Node:} In this setting, each worker node computes only the multiplication of two matrices of dimensions $\frac{r}{m}\times\frac{s}{p}$ and $\frac{s}{p}\times\frac{t}{n}$.

\textbf{Communication Load:} The master node only sends two matrices of dimensions $\frac{r}{m}\times\frac{s}{p}$ and $\frac{s}{p}\times\frac{t}{n}$ to each worker node and it receives one matrix of size $\frac{r}{m}\times\frac{t}{n}$ from worker nodes. Recall that, using CodedSketch the approximation of matrix multiplication is computed in one iteration.
 \begin{remark} \label{remark5}
In \cite{sketch}, the idea of OverSketch for matrix multiplication is proposed in which some extra count-sketches are used, as a redundancy, to mitigate the effect of stragglers. In particular,  consider two input matrices $\mathbf{A}\in\mathbb{R}^{r\times s}$ and $\mathbf{B}\in\mathbb{R}^{s\times t}$. The count-sketches of two input matrices are computed using sketch matrix $\mathbf{S}\in \mathbb{R}^{s\times z}$, as $\check{\mathbf{A}}=\mathbf{A}\mathbf{S}$ and $\check{\mathbf{B}}=\mathbf{S}^T\mathbf{B}$.
Then the block sub-matrices are created by partitioning  $\check{\mathbf{A}}$ into $\frac{rz}{q^2}$ sub-matrices and  $\check{\mathbf{B}}$ into $\frac{zt}{q^2}$ sub-matrices of size $q\times q$.  Accordingly, matrix $\mathbf{C}=\mathbf{A}\mathbf{B}$ is partitioned into $\frac{rt}{q^2}$ sub-matrices of size $q\times q$. Then each block of matrix $\mathbf{C}$ is approximately calculated by multiplying the corresponding row-block of $\check{\mathbf{A}}$ and column-block of $\check{\mathbf{B}}$.
In that scheme, the set of worker nodes is partitioned into $\eta$ subsets as $\mathcal{W}=\{\mathcal{W}_1,\dots,\mathcal{W}_{\frac{rt}{q^2}} \}$. Each subset of worker nodes computes one block of $\mathbf{C}$ approximately, where each worker node multiplies one sub-matrix in the row-block of $\check{\mathbf{A}}$ into the corresponding sub-matrix in the column-block of $\check{\mathbf{B}}$ and then the results are added.
In OverSketch, it is set $z=z^*$ for some $z^*\in\mathbb{N}$,  such that each partition $\mathcal{W}_i$ with at least $\frac{z^*}{q}$ worker nodes can approximately calculate each block of ${\mathbf{C}}$ within the following  accuracy
\begin{align}\label{1}
\mathbf{P}\bigg[||\mathbf{C}-\tilde{\mathbf{C}}_{\text{os}}||_F^2\ge \xi||\mathbf{A}||_F^2||\mathbf{B}||_F^2\bigg]\le \theta,
\end{align}
where ${\tilde{\mathbf{C}}_\text{os}}$ is the approximation of $\mathbf{C}$ using OverSketch. In \cite{sketch}, it is shown that $z^{*}=\frac{2}{\xi\theta}$.  We note that each
partition of worker nodes is dedicated to approximately calculate one block of $\mathbf{C}$, and does not help in calculating other blocks of $\mathbf{C}$. \\ 
In OverSketch, if we want the system to be resistant against $e$ stragglers for some $e\in\mathbb{N}$, then  we increase the number of worker nodes in each partition from $\frac{z^*}{q}$ to $\frac{z^*}{q}+e$. Therefore, $z=z^*+eq$.    
As a result, to calculate all blocks of ${\mathbf{C}}$, we need
\begin{equation}\label{N_over}
N_{\textrm{os}}=mn(\frac{z^*}{q}+e),
\end{equation}
total worker nodes in OverSketch, where $m=\frac{r}{q}$, $n=\frac{t}{q}$ and $z^{*}=\frac{2}{\xi\theta}$.\\   
In CodedSketch, we combine the coding theory and randomized approximation algorithms to approximately calculate the  matrix multiplication. In the proposed scheme, $d$ count-sketches of the blocks of matrix $\mathbf{C}$ will be formed after the decoding step, and the master node will recover an approximated version of $\mathbf{C}$ from these count-sketches.  In CodedSketch, according to Theorem 1, if we choose the length of the count-sketches as $b'=\mathcal{O}(\frac{1}{\epsilon^2})$ and the number of count-sketches as $d=\mathcal{O}(1/\delta)$, then we have \eqref{epsilon-delta} as the precision guarantee. 

In CodedSketch, according to \eqref{eq56}, the total number of worker nodes is computed as follows
\begin{align}\label{N_cs}
N_{cs}=(2pb'-1)(2d-1)+e.
\end{align}
Now, we compare OverSketch and CodedSketch  in terms of recovery threshold and in terms of approximation precision bounds as follows.\\
\textbf{Comparison in terms of recovery threshold:}
Here we compute the recovery threshold of OverSketch, $N_{\textrm{os}}^{th}$. Assume that out of $N_{\textrm{os}}$ in \eqref{1}, $e+1$ of them do not respond. Then it may happen that, this $e+1$ worker nodes are from one partition of worker nodes assigned to compute one specific block. Then, there are only $\frac{z^*}{q}-1$ worker nodes left for that partition, which are not enough to calculate that block within the required accuracy. 	
Therefore, the recovery threshold of OverSketch can be written as 
\begin{equation}\label{over1}
N_{\textrm{os}}^{th}=(mn-1)(\frac{z^*}{q}+e)+\frac{z^*}{q}.
\end{equation}
By eliminating $e$ from \eqref{N_over} and \eqref{over1},  	the recovery threshold is equal to
\begin{equation}
N_{\textrm{os}}^{th}=N_{\textrm{os}}-\frac{N_{\textrm{os}}}{mn}+\frac{z^*}{q}.
\end{equation}
Note that $N_{\textrm{os}}^{th}$ is a function of $N_{\textrm{os}}$, and the reason is that in OverSketch, the gain of coding is not exploited in calculating different blocks.
This is in contrary with what we have in CodedSketch. In CodedSketch, calculation in each server can help the calculation of all blocks, as we expect from coding. 
In CodedSketch method the approximated result can be calculated if any
\begin{align}
N_{\textrm{cs}}^{th}=(2pb'-1)(2d-1),
\end{align}
subset of worker nodes respond.\\
Now, we compare OverSketch and CodedSketch  with the same approximation precision bounds.
\\ 	   
\textbf{Comparison in terms of computation precision:}
Inequalities in \eqref{1} and \eqref{epsilon-delta} do not provide the precision guarantee with the same level and can not be compared. To resolve this issue, we take the following steps.\\
In CodedSketch, we have \eqref{epsilon-delta} as the precision guarantee. From \eqref{epsilon-delta} we have 
\begin{align}\label{10}
\mathbf{P}\bigg[||\mathbf{C}-\tilde{\mathbf{C}}||_{\max}\ge{\epsilon}||\mathbf{C}||_{F,(i'_m,j'_m)} \bigg]\le\delta,
\end{align}
where $i'_m=i_m \mod \frac{r}{m}$,  $j'_m=j_m \mod\frac{t}{n}$, and $(i_m,j_m)={\arg\max}_{i,j}\big\{ |[\mathbf{C}]_{i,j}-[\tilde{\mathbf{C}}]_{i,j}|\big\}$. To be able to compare OverSketch and CodedSketch, we loosen bound \eqref{10} to form the following bound
\begin{align}\label{CSbound}
\mathbf{P}\bigg[||\mathbf{C}-\tilde{\mathbf{C}}||_{F}^2\ge{rt\epsilon^2}||\mathbf{C}||^2_{F,(i'_m,j'_m)} \bigg]\le\delta.
\end{align}
According to \eqref{1} and \eqref{CSbound}, to compare the total number of worker nodes in these two schemes, let 
\begin{align}
\xi{||\mathbf{A}||_F^2||\mathbf{B}||_F^2}&=rt\epsilon^2||\mathbf{C}||^2_{F,(i'_m,j'_m)},\\
\theta&=\delta.
\end{align} 
From \eqref{N_over} and \eqref{N_cs}, the total number of worker nodes in two schemes are computed as follows
\begin{align}
N_{os}&=mn(\frac{2}{\xi\theta q}+e),\label{5}\\
N_{cs}&=(2pb'-1)(2d-1)+e=(2p\lceil{\frac{3}{\epsilon^2}}\rceil-1)(2d-1)+e\label{6},
\end{align}
where $\theta=\delta$ and ${{\xi=\frac{rt\epsilon^2||\mathbf{C}||^2_{F,(i'_m,j'_m)}}{||\mathbf{A}||_F^2||\mathbf{B}||_F^2}}}$ in \eqref{5}. \\
Now we are ready to compare these two schemes under the same level of computation precision.  Figure \ref{fig6} shows the comparison between the total number of worker nodes versus the number of stragglers in CodedSketch and OverSketch. In this simulation, for Fig.~\ref{fig6}(a) we generate two random matrices $\mathbf{A}\in\mathbb{R}^{700\times100}$ and $\mathbf{B}\in\mathbb{R}^{100\times700}$ with the elements chosen uniformly at random on the $[-100,100]$. Also, we choose $m=n=70$, $p=10$ and $\epsilon=0.1$, $\delta=0.02$. Also, for Fig.~\ref{fig6}(b) we generate two random matrices $\mathbf{A}\in\mathbb{R}^{500\times50}$ and $\mathbf{B}\in\mathbb{R}^{50\times500}$ with the elements chosen uniformly at random on the $[-100,100]$. Also, we choose $m=n=50$, $p=5$ and $\epsilon=0.1$, $\delta=0.2$.  Figure \ref{fig6} shows that compared to OverSketch, in some cases, CodedSketch can reduce the total number of worker nodes by order of magnitudes with the same approximation precision bound. It is worth noting that to make the precision guarantee bound similar for two schemes, we took some steps to form \eqref{CSbound}, which is a very loose bound of CodedSketch but still, we can see that these two schemes are comparable.
\begin{figure}[htp]
\centering
\begin{tabular}{@{}c@{}c}
	\includegraphics[scale=0.55,draft=false]{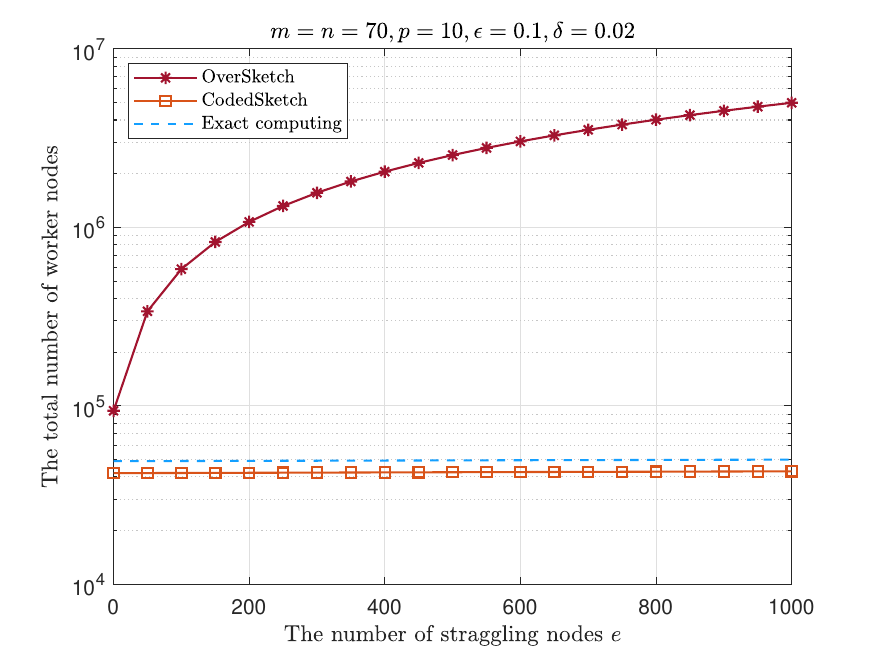}&
	\includegraphics[scale=0.55,draft=false]{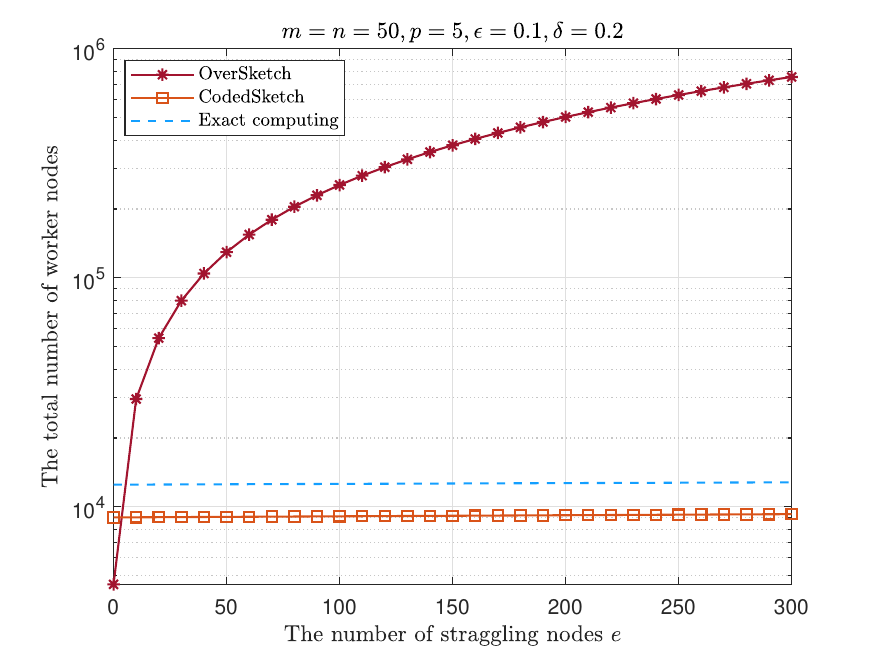}\\
	(a)& (b)
\end{tabular}
\caption{Comparison between CodedSketch and OverSketch in the total number of worker nodes versus the number of stragglers.}
\label{fig6}
\end{figure} 
 \end{remark} 	
  \section{Conclusion}
In this paper, we proposed a coded sketch scheme to approximately compute the multiplication of two large-scale matrices in a distributed straggler-resistant system. 
The proposed scheme exploits the structure of matrices in multiplication, and still uses the randomized pre-compression scheme based on sketching schemes. 
This scheme provides  an upper-bound on the recovery threshold as a function of the required accuracy of computation and the probability that the required accuracy can be violated. In addition, it provides an upper-bound on the recovery threshold for the case that the result of the multiplication is sparse, and the exact result is required.  
The next step would be to consider more general functions, specially polynomial functions.
\section{Acknowledgment}
This research was supported by Iran National Science Foundation (INSF) under contract No. 98019168. 
 	

\bibliography{References}
\bibliographystyle{ieeetr}
\appendix

\subsection{Proof of \eqref{57} }\label{appC}

	In \eqref{eq54} and \eqref{eq55} it can be seen that $\mathbf{F}(x,\alpha,\eta)=\mathbf{F}_\eta(x,\alpha)$ and $\mathbf{G}(x,\alpha,\eta)=\mathbf{G}_\eta(x,\alpha)$ for $\eta=1,\dots ,d$, and we know that each $\mathbf{F}_\eta(x,\alpha)\mathbf{G}_\eta(x,\alpha)^T$ is a sketch polynomial for $\hat{\mathbf{C}}(x)$, i.e.,
	\begin{equation}\label{38}
	\mathbf{P}_{\eta,\eta}(x,\alpha) \buildrel \Delta \over =\mathbf{F}_\eta(x,\alpha)\mathbf{G}_\eta(x,\alpha)^T=\sum_{i=0}^{m-1}\sum_{j=0}^{n-1}s_{\eta}(i){\tilde{s}_{\eta}(j)\hat{\mathbf{A}}_{i,0}(x)\hat{\mathbf{B}}_{0,j}(x)^T\alpha^{h_\eta(i)+\tilde{h}_\eta(j)}}.
	\end{equation}
	Also, \eqref{38} can be written as follows
	\begin{equation}\label{39}
	\mathbf{P}_{\eta,\eta}(x,\alpha)=\sum_{k=0}^{2b'-2}{\mathbf{P}_k^{(\eta,\eta)}(x)\alpha^k},
	\end{equation} 
	where in this expansion,
	\begin{equation}\label{71}
	\mathbf{P}_k^{(\eta,\eta)}(x)=\hspace{-0.5cm}\sum_{i,j :\hspace{0.1cm} h_\eta(i)+\tilde{h}_\eta(j)=k}\hspace{-0.5cm}{s_\eta(i)\tilde{s}_\eta(j)}\hat{\mathbf{A}}_{i,0}(x)\hat{\mathbf{B}}_{0,j}(x)^T=\hspace{-0.5cm}\sum_{i,j :\hspace{0.1cm}h_\eta(i)+\tilde{h}_\eta(j)=k}\hspace{-0.5cm}{s_\eta(i)\tilde{s}_\eta(j)\hat{\mathbf{C}}_{i,j}(x)}.
	\end{equation}
	As mentioned before, the block $(i,j)$ of $\hat{\mathbf{C}}(x)$, i.e., $\hat{\mathbf{C}}_{i,j}(x)$ is a polynomial of degree $(2p-2)$ where we can recover the linear combination of some blocks of $\mathbf{C}$ as the coefficient of $x^{p-1}$. So, in the sketch polynomial of $\hat{\mathbf{C}}(x)$ there is a hidden count-sketch for blocks of $\mathbf{C}$. Also, \eqref{39} can be explained in the following form
	\begin{equation}\label{41}
	\mathbf{P}_{\eta,\eta}(x,\alpha)=\sum_{k=0}^{2b'-2}\sum_{k'=0}^{2p-2}{\mathbf{P}_{k,k'}^{(\eta,\eta)}x^{k'}\alpha^k},
	\end{equation}
	where the linear combinations of some blocks of $\mathbf{C}$ are located in $\mathbf{P}_{k,p-1}^{(\eta,\eta)}$ for $k=0,..,2b'-2$. Thus, according to \eqref{71}, and \eqref{41}, we have
	\begin{equation}\label{42}
	\mathbf{P}_{k,p-1}^{(\eta,\eta)}=\hspace{-0.5cm}\sum_{i,j: h_\eta(i)+\tilde{h}_\eta(j)=k}\hspace{-0.5cm}{s_\eta(i)\tilde{s}_\eta(j)\mathbf{C}_{i,j}},
	\end{equation}
	where $\mathbf{C}_{i,j}$ is the block $(i,j)$ of $\mathbf{C}$.

\end{document}